\documentclass[english]{IEEEtran}
\usepackage[T1]{fontenc}
\usepackage[latin9]{inputenc}
\usepackage{array}
\usepackage{multirow}
\usepackage{amsmath}
\usepackage{amsthm}
\usepackage{amssymb}
\usepackage{graphicx}

\makeatletter

\providecommand{\tabularnewline}{\\}

  \theoremstyle{definition}
  \newtheorem{defn}{\protect\definitionname}
  \theoremstyle{plain}
  \newtheorem{lem}{\protect\lemmaname}
  \theoremstyle{remark}
  \newtheorem{rem}{\protect\remarkname}
  \theoremstyle{plain}
  \newtheorem{thm}{\protect\theoremname}
  \theoremstyle{plain}
  \newtheorem{prop}{\protect\propositionname}
  \theoremstyle{plain}
  \newtheorem{cor}{\protect\corollaryname}

\usepackage{mathptmx}

\makeatother

\usepackage{babel}
\providecommand{\corollaryname}{Corollary}
\providecommand{\definitionname}{Definition}
\providecommand{\lemmaname}{Lemma}
\providecommand{\propositionname}{Proposition}
\providecommand{\remarkname}{Remark}
\providecommand{\theoremname}{Theorem}

\begin{document}

\title{Modeling and Analysis of Leaky Deception using Signaling Games with
Evidence}

\author{Jeffrey Pawlick, Edward Colbert, and Quanyan Zhu}
\maketitle
\begin{abstract}
Deception plays critical roles in economics and technology, especially
in emerging interactions in cyberspace. Holistic models of deception
are needed in order to analyze interactions and to design mechanisms
that improve them. Game theory provides such models. In particular,
existing work models deception using signaling games. But signaling
games inherently model deception that is undetectable. In this paper,
we extend signaling games by including a detector that gives off probabilistic
warnings when the sender acts deceptively. Then we derive pooling
and partially-separating equilibria of the game. We find that 1) high-quality
detectors eliminate some pure-strategy equilibria, 2) detectors with
high true-positive rates encourage more honest signaling than detectors
with low false-positive rates, 3) receivers obtain optimal outcomes
for equal-error-rate detectors, and 4) surprisingly, deceptive senders
sometimes benefit from highly accurate deception detectors. We illustrate
these results with an application to defensive deception for network
security. Our results provide a quantitative and rigorous analysis
of the fundamental aspects of detectable deception.\end{abstract}


\section{Introduction\label{sec:intro}}

Deception is a fundamental facet of interactions ranging from biology
\cite{cott1940adaptive} to criminology \cite{vrij2008increasing}
and from economics \cite{gneezy2005deception} to the Internet of
Things (IoT) \cite{pawlick2017trust}. Cyberspace creates particular
opportunities for deception, since information lacks permanence, imputing
responsibility is difficult \cite{janczewski2008cyber}, and some
agents lack repeated interactions \cite{minhas2011multifaceted}.
For instance, online interactions are vulnerable to identify theft
 and spear phishing, and authentication in the IoT suffers from
a lack of infrastructure and local computational resources \cite{atzori2010internet}. 

Defenders also implement deception. Traditional security approaches
such as firewalls and role-based access control (RBAC) have proved
insufficient against insider attacks and advanced persistent threats
(APTs). Hence, defenders in the security and privacy domains have
proposed, \emph{e.g.}, honeynets \cite{carroll2011game}, moving target
defense \cite{zhu2013game},  and mix networks \cite{zhang2010gpath}.
Using these techniques, defenders can obscure valuable information
such as personally identifiable information or the configuration of
a network. They can also send false information to attackers in order
to waste their resources or attract them away from critical assets.
Both malicious and defensive deception have innumerable implications
for cybersecurity.

\subsection{Quantifying Deception using Signaling Games}

Modeling deceptive interactions online and in the IoT would allow
government policymakers, technological entrepreneurs, and vendors
of cyber-insurance to predict changes in these interactions due to
legislation, new technology, or risk mitigation. While deception is
studied in each of these domains individually, a general, quantitative,
and systematic understanding of deception seems to be lacking. 

What commonalities underlie all forms of deception? Deception is 1)
information asymmetric, 2) dynamic, and 3) strategic. In deceptive
interactions, one party (hereafter, the \emph{sender}) possesses private
information unknown to the other party (hereafter, the \emph{receiver}).
Based on her private information, the sender communicates a possibly-untruthful
message to the receiver. Then the receiver forms a belief about the
private information of the sender, and chooses an action. The players
act \emph{strategically}, in the sense that they each seek a result
that corresponds to their individual incentives. 

Non-cooperative game theory provides a set of tools to study interactions
between multiple, strategic agents. In the equilibrium of a game,
agents adapt their strategies to counter the strategies of the other
agents. This rationality models the sophisticated behavior characteristic
of deception. In particular, \emph{cheap-talk signaling games} \cite{crawford1982strategic}
model interactions that are strategic, dynamic, and information-asymmetric.
In cheap-talk signaling games, a sender $S$ with private information
communicates a message to a receiver $R,$ who acts upon it. Then
both players receive utility based on the private information of $S$
and the action of $R.$ Recently, signaling games have been used to
model deceptive interactions in resource allocation \cite{zhuang2010modeling},
network defense \cite{carroll2011game,pawlick2015deception}, and
cyber-physical systems \cite{pawlick2015flip}.

\subsection{Cost and Detection in Signaling Games}

The phrase \emph{cheap talk} signifies that the utility of both players
is independent of the message that $S$ communicates to $R.$ In cheap-talk
signaling games, therefore, there is no cost or risk of lying \emph{per
se}. Truth-telling sometimes emerges in equilibrium, but not through
the penalization or detection of lying. We can say that cheap-talk
signaling games model deception that is \emph{undetectable} and \emph{costless.}

In economics literature, Navin Kartik has proposed a signaling game
model that rejects the second of these two assumptions \cite{kartik2009strategic}.
In Kartik's model, $S$ pays an explicit cost to send a message that
does not truthfully represent her private information. This cost could
represent the effort required, \emph{e.g.}, to obfuscate data, to
suppress a revealing signal, or to fabricate misleading data. In equilibrium,
the degree of deception depends on the lying cost. Contrary to cheap-talk
games, Kartik's model studies deception that has a cost. Yet the deception
is still undetectable\emph{.}

In many scenarios, however, deception can be detected with some probability.
Consider the issue of so-called \emph{fake news} in social media.
Fake news stories about the 2016 U.S. Presidential Election reportedly
received more engagement on Facebook than news from real media outlets
\cite{buzzfeed2016fake}. In the wake of the election, Facebook announced
its own program to detect fake news and alert users about suspicious
articles. As another example, consider deceptive product reviews in
online marketplaces. Linguistic analysis has been used to detect this
\emph{deceptive opinion spam} \cite{ott2011finding}. Finally, consider
deployment of honeypots as a technology for defensive deception. Attackers
have developed tools that detect the virtual machines often used to
host honeypots.

Therefore, we propose a model of signaling games in which a detector
emits probabilistic evidence of deception. The detector can be interpreted
in two equally valid ways. It can be understood as a technology that
$R$ uses to detect deception, such as a phishing detector in an email
client. Alternatively, it can be understood as the inherent tendency
of $S$ to emit cues to deception when she misrepresents her private
information. For instance, in interpersonal deception, lying is cognitively
demanding, and sometimes liars give off cues from these cognitive
processes \cite{vrij2008increasing}. $R$ uses the evidence from
the detector to form a belief about whether $S$'s message honestly
conveys her private information.

Naturally, several questions arise. What is the equilibrium of the
game? How, if at all, is the equilibrium different than that of a
traditional cheap-talk signaling game? Is detection always harmful
for the sender? Finally, how does the design of the detector affect
the equilibrium? Our paper addresses each of these questions.

\subsection{Contributions\label{sub:contributions}}

We present the following contributions:
\begin{enumerate}
\item We describe a model for deception based on cheap-talk signaling games
with a probabilistic deception detector.
\item We find all pure-strategy equilibria of the game, and we derive mixed-strategy
equilibria in regimes that do not support pure-strategy equilibria.
\item We prove that detectors that prioritize high true-positive rates encourage
more honest equilibrium behavior than detectors that prioritize low
false-positive rates.
\item Numerical results suggest that the sender (deceptive agent) benefits
from an accurate detector in some parameter regimes. The receiver
(agent being deceived) benefits from an accurate detector, and preferably
one with an equal error rate.
\item We apply our analytical results to a case study in defensive deception
for network security.
\end{enumerate}

\subsection{Related Work\label{sub:Related-Work}}

Signaling games with evidence are related to \emph{hypothesis testing},
\emph{inspection games}, and \emph{trust management}. Hypothesis testing
evaluates the truthfulness of claims based on probabilistic evidence
\cite{levy2008principles}. Inspection games embed a hypothesis testing
problem inside of a two-player game \cite{avenhaus2002inspection}.
An \emph{inspector} designs an inspection technique in order to motivate
an \emph{inspectee} to follow a rule or regulation. The inspector
chooses a probability with which to execute an inspection, and chooses
whether to trust the inspectee based on the result. Our work adds
the concept of signaling to the framework of inspection games. Finally,
our model of deception can be seen as a dual to models for trust management
\cite{minhas2011multifaceted}, which quantify technical and behavioral
influences on the transparency and reliability of agents in distributed
systems.

Economics literature includes several classic contributions to signaling
games. Crawford and Sobel's seminal paper is the foundation of cheap-talk
signaling games \cite{crawford1982strategic}. In this paper, a sender
can just as easily misrepresent her private information as truthfully
represent it. From the opposite vantage point, models from Milgrom
\cite{milgrom1981good}, Grossman \cite{grossman1981informational},
and Grossman and Heart \cite{grossman1980disclosure} study games
of verifiable disclosure. In these models, a sender can choose to
remain silent or to disclose information. But if the sender chooses
to disclose information, then she must do so truthfully. 

One way of unifying cheap-talk games and games of verifiable disclosure
is to assign an explicit cost to misrepresenting the truth. This idea
is due to Kartik \cite{kartik2009strategic}. Cheap-talk games are
a special case of this model in which the cost of lying zero, and
games of verifiable disclosure are a special case in which the cost
is infinite. Our model also bridges cheap-talk games and games of
verifiable disclosure. But we do this using detection rather than
lying cost. Cheap-talk games are a special case in which the detector
gives alarms randomly, and games of verifiable disclosure are a special
case in which the detector is perfect.

The rest of the paper proceeds as follows. Section \ref{sec:model}
describes our model and equilibrium concept. In Section \ref{sec:eqResults},
we find pure-strategy and mixed-strategy equilibria. Then Section
\ref{sec:compStatics} evaluates the sensitivity of the equilibria
to changes in detector characteristics. Section \ref{sec:app} describes
the application. Finally, we discuss the implications of our results
in Section \ref{sec:discuss}.

\section{Model\label{sec:model}}

\begin{figure}
\begin{centering}
\includegraphics[width=0.9\columnwidth]{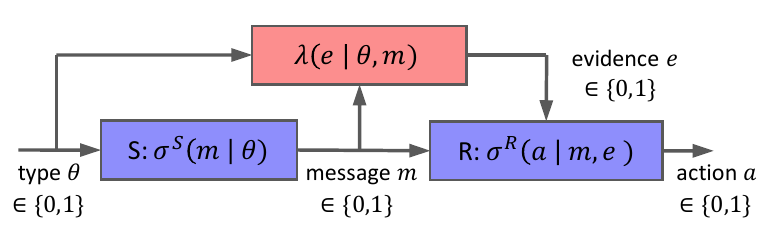}
\par\end{centering}

\caption{\label{fig:sigWEv}Signaling games with evidence add the red detector
block to the $S$ and $R$ blocks. The probability $\lambda(e\,|\,\theta,m)$
of emitting evidence $e$ depends on $S$'s type $\theta$ and the
message $m$ that she transmits. }
\end{figure}

Signaling games are two-player games between a sender ($S$) and a
receiver ($R$). These games are \emph{information asymmetric}, because
$S$ possesses information that is unknown to $R.$ They are also
\emph{dynamic}, because the players' actions are not simultaneous.
$S$ transmits a message to $R,$ and then $R$ acts upon the message.
Generally, signaling games can be \emph{non-zero-sum}, which means
that the objectives of $S$ and $R$ are not direct negations of each
other's objectives. Figure \ref{fig:sigWEv} depicts the traditional
signaling game between $S$ and $R,$ augmented by a detector block.
We call this augmented signaling game a \emph{signaling game with
evidence}.

\subsection{Types, Messages, Evidence, Actions, and Beliefs}

\begin{table}
\caption{\label{tab:notation}Summary of Notation}

\centering{}%
\begin{tabular}{|c|c|}
\hline 
Notation & Meaning\tabularnewline
\hline 
\hline 
$S,\:R$  & Sender and Receiver\tabularnewline
\hline 
$\theta\in\Theta,\:m\in M,\:a\in A$ & Types, Messages, Actions\tabularnewline
\hline 
$u^{X}\left(\theta,m,a\right)$ & Utility Functions of Player $X\in\left\{ S,R\right\} $\tabularnewline
\hline 
$p\left(\theta\right)$ & Prior Probability of $\theta$\tabularnewline
\hline 
$\sigma^{S}\left(m\,|\,\theta\right)$ & Mixed Strategy of $S$ of Type $\theta$\tabularnewline
\hline 
$\lambda\left(e\,|\,\theta,m\right)$ & Probability of $e$ for Type $\theta$ \& Message $m$\tabularnewline
\hline 
$\sigma^{R}\left(a\,|\,m,e\right)$ & Mixed Strategy of $R$ given $m,\,e$ \tabularnewline
\hline 
$\mu^{R}\left(\theta\,|\,m,e\right)$ & Belief of $R$ that $S$ is of Type $\theta$\tabularnewline
\hline 
$\bar{u}^{S}\left(\sigma^{S},\sigma^{R}\,|\,\theta\right)$ & Expected Utility for $S$ of Type $\theta$\tabularnewline
\hline 
$\bar{u}^{R}\left(\sigma^{R}\,|\,\theta,m,e\right)$ & Expected Utility for $R$ given $\theta,$ $m,$ $e$\tabularnewline
\hline 
$\alpha\in\left[0,1\right]$ & ``Size'' of Detector (False Positive Rate)\tabularnewline
\hline 
$\beta\in$$\left[\alpha,1\right]$ & ``Power'' of Detector (True Positive Rate)\tabularnewline
\hline 
\end{tabular}
\end{table}
We consider binary information and action spaces in order to simplify
analysis\footnote{Future work can consider continuous spaces for each quantity, perhaps
building upon the  continuous space model due to Crawford and Sobel
\cite{crawford1982strategic}.}. Table \ref{tab:notation} summarizes the notation. Let $\theta\in\Theta=\{0,1\}$
denote the private information of $S.$ Signaling games refer to $\theta$
as a \emph{type}. The type could represent \emph{e.g.}, whether the
sender is a malicious or benign actor, whether she has one set of
preferences over another, or whether a given event observable to $S$
but not to $R$ has occurred. The type is drawn\footnote{Harsanyi conceptualized type selection as a randomized move by a non-strategic
player called \emph{nature }(in order to map an incomplete information
game to one of complete information) \cite{harsanyi1967games}.} from a probability distribution $p,$ where $\sum_{\theta\in\Theta}p\left(\theta\right)=1$
and $\forall\theta\in\Theta,$ $p(\theta)\geq0.$

Based on $\theta$, $S$ chooses a message $m\in M=\left\{ 0,1\right\} .$
She may use mixed strategies, \emph{i.e.}, she may select each $m$
with some probability. Let $\sigma^{S}\in\Gamma^{S}$ denote the strategy
of $S,$ such that $\sigma^{S}(m\,|\,\theta)$ gives the probability
with which $S$ sends message $m$ given that she is of type $\theta.$
The space of strategies satisfies
\[
\Gamma^{S}=\biggl\{\sigma^{S}\,|\,\forall\theta,\,\underset{m\in M}{\sum}\sigma^{S}\left(m\,|\,\theta\right)=1;\:\forall\theta,m,\;\sigma^{S}\left(m\,|\,\theta\right)\geq0\biggr\}.
\]
Since $\Theta$ and $M$ are identical, a natural interpretation of
an honest message is $m=\theta,$ while a deceptive message is represented
by $m\neq\theta.$ 

Next, the detector emits evidence based on whether the message $m$
is equal to the type $\theta.$ The detector emits $e\in E=\left\{ 0,1\right\} $
by the probability $\lambda\left(e\,|\,\theta,m\right).$ Let $e=1$
denote an \emph{alarm} and $e=0$ \emph{no alarm}. The probability
with which a detector records a true positive is called the \emph{power}
$\beta\in[0,1]$ of the detector. For simplicity, we set both true-positive
rates to be equal: $\beta=\lambda(1\,|\,0,1)=\lambda(1\,|\,1,0).$
Similarly, let $\alpha$ denote the \emph{size }of the detector, which
refers to the false-positive rate. We have $\alpha=\lambda(1\,|\,0,0)=\lambda(1\,|\,1,1).$
A valid detector has $\beta\geq\alpha.$ This is without loss of generality,
because otherwise $\alpha$ and $\beta$ can be relabeled.

After receiving both $m$ and $e,$ $R$ chooses an action $a\in A=\left\{ 0,1\right\} .$
$R$ may also use mixed strategies. Let $\sigma^{R}\in\Gamma^{R}$
denote the strategy of $R$ such that his mixed-strategy probability
of playing action $a$ given message $m$ and evidence $e$ is $\sigma^{R}(a\,|\,m,e).$
The space of strategies is $\Gamma^{R}=$
\[
\Bigl\{\sigma^{R}\,|\,\forall m,e,\:\underset{a\in A}{\sum}\sigma^{R}\left(a\,|\,m,e\right)=1;\,\forall e,m,a,\;\sigma^{R}\left(a\,|\,m,e\right)\geq0\Bigr\}.
\]

Based on $m$ and $e,$ $R$ forms a belief\footnote{The \emph{Stanford Encyclopedia of Philosophy} lists among several
definitions of deception: ``to intentionally cause to have a false
belief that is known and believed to be false'' \cite{stanford2016deception}.
This suggests that belief formation is an important aspect of deception. } about the type $\theta$ of $S.$ For all $\theta,$ $m,$ and $e,$
define $\mu^{R}:\,\Theta\to\left[0,1\right]$ such that $\mu^{R}(\theta\,|\,m,e)$
gives the likelihood with which $R$ believes that $S$ is of type
$\theta$ given message $m$ and evidence $e.$ $R$ uses belief $\mu^{R}$
to decide which action to chose.

\subsection{Utility Functions}

Let $u^{S}:\,\Theta\times M\times A\to\mathbb{R}$ denote a utility
function for $S$ such that $u^{S}\left(\theta,m,a\right)$ gives
the utility that she receives when her type is $\theta,$ she sends
message $m,$ and $R$ plays action $a.$ Similarly, let $u^{R}:\,\Theta\times M\times A$
denote $R$'s utility function so that $u^{R}\left(\theta,m,a\right)$
gives his payoff under the same scenario. 

Only a few assumptions are necessary to characterize a deceptive interaction.
Assumption $1$ is that $u^{S}$ and $u^{R}$ do not depend (exogenously)
on $m,$ \emph{i.e.}, the interaction is a cheap-talk game. Assumptions
2-3 state that $R$ receives higher utility if he correctly chooses
$a=\theta$ than if he chooses $a\neq\theta.$ Formally, 
\[
u^{R}\left(0,m,0\right)>u^{R}\left(0,m,1\right),\;u^{R}\left(1,m,0\right)<u^{R}\left(1,m,1\right).
\]
Finally, Assumptions 4-5 state that $S$ receives higher utility if
$R$ chooses $a\neq\theta$ than if he chooses $a=\theta.$ That is,
\[
u^{S}\left(0,m,0\right)<u^{S}\left(0,m,1\right),\;u^{S}\left(1,m,0\right)>u^{S}\left(1,m,1\right).
\]
Together, Assumptions 1-5 characterize a \emph{cheap-talk signaling
game with evidence}.

Define an expected utility function $\bar{u}^{S}:\,\Gamma^{S}\times\Gamma^{R}\to\mathbb{R}$
such that $\bar{u}^{S}\left(\sigma^{S},\sigma^{R}\,|\,\theta\right)$
gives the expected utility to $S$ when she plays strategy $\sigma^{S},$
given that she is of type $\theta.$ This expected utility is given
by
\begin{multline*}
\bar{u}^{S}\left(\sigma^{S},\sigma^{R}\,|\,\theta\right)=\sum_{a\in A}\sum_{e\in E}\sum_{m\in M}\\
\sigma^{R}\left(a\,|\,m,e\right)\lambda\left(e\,|\,\theta,m\right)\sigma^{S}\left(m\,|\,\theta\right)u^{S}\left(\theta,m,a\right).
\end{multline*}
Next define $\bar{u}^{R}:\,\Gamma^{R}\to\mathbb{R}$ such that $\bar{u}^{R}(\sigma^{R}\,|\,\theta,m,e)$
gives the expected utility to $R$ when he plays strategy $\sigma^{R}$
given message $m,$ evidence $e,$ and sender type $\theta.$ The
expected utility function is given by
\[
\bar{u}^{R}(\sigma^{R}\,|\,\theta,m,e)=\sum_{a\in A}\sigma^{R}\left(a\,|\,m,e\right)u^{R}\left(\theta,m,a\right).
\]

\subsection{Equilibrium Concept}

In two-player games, Nash equilibrium defines a strategy profile in
which each player best responds to the optimal strategies of the other
player \cite{nash1950equilibrium}. Signaling games motivate the extension
of Nash equilibrium in two ways. First, information asymmetry requires
$R$ to maximize his expected utility over the possible types of $S.$
An equilibrium in which $S$ and $R$ best respond to each other's
strategies given some belief $\mu^{R}$ is called a \emph{Bayesian
Nash equilibrium }\cite{harsanyi1967games}. We also require $R$
to update $\mu^{R}$ rationally. \emph{Perfect Bayesian Nash equilibrium}
(PBNE) captures this constraint. Definition \ref{def:PBNE} applies
PBNE to our game.
\begin{defn}
\label{def:PBNE}(Perfect Bayesian Nash Equilibrium \cite{fudenberg1991game})
A PBNE of a cheap-talk signaling game with evidence is a strategy
profile $(\sigma^{S*},\sigma^{R*})$ and posterior beliefs $\mu^{R}(\theta\,|\,m,e)$
such that
\begin{equation}
\forall\theta\in\Theta,\:\sigma^{S*}\in\underset{\sigma^{S}\in\Gamma^{S}}{\arg\max}\,\bar{u}^{S}\left(\sigma^{S},\sigma^{R*}\,|\,\theta\right),\label{eq:senderOpt}
\end{equation}
$\forall m\in M,\:\forall e\in E,$
\begin{equation}
\sigma^{R*}\in\underset{\sigma^{R}\in\Gamma^{R}}{\arg\max}\,\sum_{\theta\in\Theta}\mu^{R}\left(\theta\,|\,m,e\right)\bar{u}^{R}(\sigma^{R}\,|\,\theta,m,e),\label{eq:receiverOpt}
\end{equation}
and if $\underset{\tilde{\theta}\in\Theta}{\sum}\lambda(e\,|\,\tilde{\theta},m)\sigma^{S}(m\,|\,\tilde{\theta})p(\tilde{\theta})>0,$
then
\begin{equation}
\mu^{R}\left(\theta\,|\,m,e\right)=\frac{\lambda\left(e\,|\,\theta,m\right)\mu^{R}\left(\theta\,|\,m\right)}{\underset{\tilde{\theta}\in\Theta}{\sum}\lambda\left(e\,|\,\tilde{\theta},m\right)\mu^{R}\left(\tilde{\theta}\,|\,m\right)},\label{eq:updateEv}
\end{equation}
where
\begin{equation}
\mu^{R}\left(\theta\,|\,m\right)=\frac{\sigma^{S}\left(m\,|\,\theta\right)p\left(\theta\right)}{\underset{\hat{\theta}\in\Theta}{\sum}\sigma^{S}\left(m\,|\,\hat{\theta}\right)p\left(\hat{\theta}\right)}.\label{eq:updateMsg}
\end{equation}
If $\underset{\tilde{\theta}\in\Theta}{\sum}\lambda(e\,|\,\tilde{\theta},m)\sigma^{S}(m\,|\,\tilde{\theta})p(\tilde{\theta})=0,$
then $\mu^{R}\left(\theta\,|\,m,e\right)$ may be set to any probability
distribution over $\Theta.$ 
\end{defn}
Equations (\ref{eq:updateEv})-(\ref{eq:updateMsg}) require the belief
to be set according to Bayes' Law. First, $R$ updates her belief
according to $m$ using Eq. (\ref{eq:updateMsg}). Then $R$ updates
her belief according to $e$ using Eq. (\ref{eq:updateEv}). This
step is not present in the traditional definition of PBNE for signaling
games. 

There are three categories of PBNE: \emph{separating}, \emph{pooling},
and \emph{partially-separating} equilibria. These are defined based
on the strategy of $S.$ In separating PBNE, the two types of $S$
transmit opposite messages. This allows $R$ to infer $S$'s type
with certainty. In pooling PBNE, both types of $S$ send messages
with identical probabilities. That is, $\forall m\in M,\:\sigma^{S}(m\,|\,0)=\sigma^{S}(m\,|\,1).$
This makes $m$ useless to $R.$ $R$ updates his belief based only
on the evidence $e.$ Equations (\ref{eq:updateEv})-(\ref{eq:updateMsg})
yield
\begin{equation}
\mu^{R}\left(\theta\,|\,m,e\right)=\frac{\lambda\left(e\,|\,\theta,m\right)p\left(\theta\right)}{\underset{\tilde{\theta}\in\Theta}{\sum}\lambda\left(e\,|\,\tilde{\theta},m\right)p\left(\tilde{\theta}\right)}.\label{eq:updatePooling}
\end{equation}

In partially-separating PBNE, the two types of $S$ transmit messages
with different, but not completely opposite, probabilities. In other
words, $\forall m\in M,$ $\sigma^{S}(m\,|\,0)\neq\sigma^{S}(m\,|\,1),$
and $\sigma^{S}(m\,|\,0)\neq1-\sigma^{S}(m\,|\,1).$ Equations (\ref{eq:updateEv})-(\ref{eq:updateMsg})
allow $R$ to update his belief, but the belief remains uncertain.

\section{Equilibrium Results\label{sec:eqResults}}

In this section, we find the PBNE of the cheap-talk signaling game
with evidence. We present the analysis in four steps. In Subsection
\ref{sub:Prior-Probability-Regimes}, we solve the optimality condition
for $R,$ which determines the structure of the results. In Subsection
\ref{sub:Optimality-Condition-for}, we solve the optimality condition
for $S,$ which determines the equilibrium beliefs $\mu^{R}.$ We
present the pooling equilibria of the game in Subsection \ref{sub:Pooling-PBNE}.
Some parameter regimes do not admit any pooling equilibria. For those
regimes, we derive partially-separating equilibria in Subsection \ref{sub:Partially-Separating-PBNE}.

First, Lemma \ref{lem:noSep} notes that one class of equilibria is
not supported.
\begin{lem}
\label{lem:noSep}Under Assumptions 1-5, the game admits no separating
PBNE. 
\end{lem}
The proof is straightforward, so we omit it here. Lemma \ref{lem:noSep}
results from the opposing utility functions of $S$ and $R.$ $S$
wants to deceive $R,$ and $R$ wants to correctly guess the type.
It is not incentive-compatible for $S$ to fully reveal the type by
choosing a separating strategy.

\subsection{Prior Probability Regimes\label{sub:Prior-Probability-Regimes}}

\begin{figure*}
\begin{centering}
\includegraphics[width=1\textwidth]{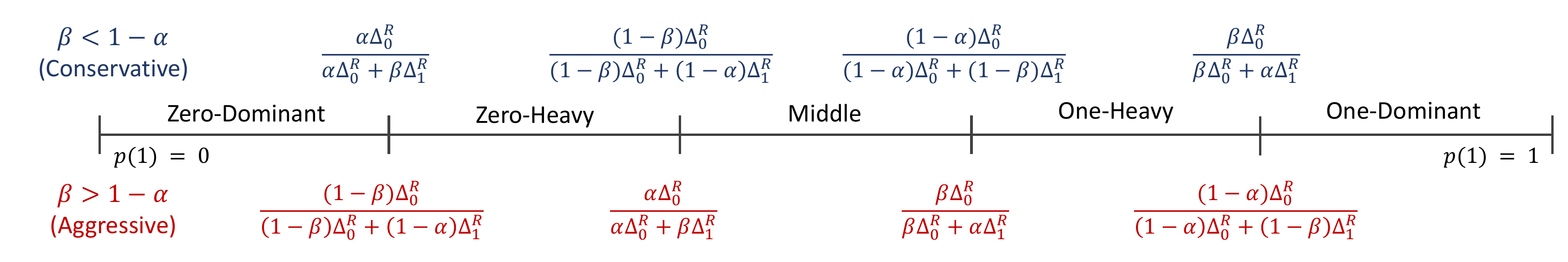}
\par\end{centering}

\caption{\label{fig:eqRegions}PBNE differ within five prior probability regimes.
In the Zero-Dominant regime, $p(\theta=1)\approx0,$ \emph{i.e.},
type $0$ dominates. In the Zero-Heavy regime, $p(\theta=1)$ is slightly
higher, but still low. In the Middle regime, the types are mixed almost
evenly. The One-Heavy regime has a higher $p(\theta=1),$ and the
One-Dominant regime has $p(\theta=1)\approx1.$ The definitions of
the regime boundaries depend on whether the detector is conservative
or aggressive.}
\end{figure*}
Next, we look for pooling PBNE. Consider $R$'s optimal strategies
$\sigma^{R*}$ in this equilibrium class. Note that if the sender
uses a pooling strategy on message $m$ (\emph{i.e.}, if $S$ with
both $\theta=0$ and $\theta=1$ send message $m$), then $\sigma^{R*}(1\,|\,m,e)$
gives $R$'s optimal action $a$ after observing evidence $e.$ Messages
do not reveal anything about the type $\theta,$ and $R$ updates
his belief using Eq. (\ref{eq:updatePooling}). For brevity, define
the following notations:
\begin{equation}
\Delta_{0}^{R}\triangleq u^{R}\left(\theta=0,m,a=0\right)-u^{R}\left(\theta=0,m,a=1\right),\label{eq:CBR0}
\end{equation}
\begin{equation}
\Delta_{1}^{R}\triangleq u^{R}\left(\theta=1,m,a=1\right)-u^{R}\left(\theta=1,m,a=0\right).\label{eq:CBR1}
\end{equation}
$\Delta_{0}^{R}$ gives the benefit to $R$ for correctly guessing
the type when $\theta=0,$ and $\Delta_{1}^{R}$ gives the benefit
to $R$ for correctly guessing the type when $\theta=1.$ Since the
game is a cheap-talk game, these benefits are independent of $m.$
Lemmas \ref{lem:BRregions}-\ref{lem:BRactions} solve for $\sigma^{R*}$
within five regimes of the prior probability $p(\theta)$ of each
type $\theta\in\{0,1\}.$ Recall that $p(\theta)$ represents $R$'s
belief that $S$ has type $\theta$ \emph{before }$R$ observes $m$
or $e.$ 
\begin{lem}
\label{lem:BRregions} For pooling PBNE, $R$'s optimal actions $\sigma^{R*}$
for evidence $e$ and messages $m$ on the equilibrium path\footnote{In pooling PBNE, the message ``on the equilibrium path'' is the
one that is sent by both types of $S.$ Messages ``off the equilibrium
path'' are never sent in equilibrium, although determining the actions
that $R$ \emph{would play} if $S$ were to transmit a message off
the path is necessary in order to determine the existence of equilibria.} vary within five regimes of $p(\theta).$ The top half of Fig. \ref{fig:eqRegions}
lists the boundaries of these regimes for detectors in which $\beta<1-\alpha,$
and the bottom half of Fig. \ref{fig:eqRegions} lists the boundaries
of these regimes for detectors in which $\beta>1-\alpha.$ \end{lem}
\begin{IEEEproof}
See Appendix \ref{ap:brR}.\end{IEEEproof}
\begin{rem}
\label{rem:classesDetectors}Boundaries of the equilibrium regimes
differ depending on the relationship between $\beta$ and $1-\alpha.$
$\beta$ is the true-positive rate and $1-\alpha$ is the true-negative
rate. Let us call detectors with $\beta<1-\alpha$ \emph{conservative
}detectors, detectors with $\beta>1-\alpha$ \emph{aggressive detectors},
and detectors with $\beta=1-\alpha$ \emph{equal-error-rate }(\emph{EER})
detectors. Aggressive detectors have high true-positive rates but
also high false-positive rates. Conservative detectors have low false-positive
rates but also low true-positive rates. Equal-error-rate detectors
have an equal rate of false-positives and false-negatives. 
\end{rem}

\begin{rem}
The regimes in Fig. \ref{fig:eqRegions} shift towards the right as
$\Delta_{0}^{R}$ increases. Intuitively, a higher $p(1)$ is necessary
to balance out the benefit to $R$ for correctly identifying a type
$\theta=0$ as $\Delta_{0}^{R}$ increases. The regimes shift towards
the left as $\Delta_{1}^{R}$ increases for the opposite reason.
\end{rem}
Lemma \ref{lem:BRactions} gives the optimal strategies of $R$ in
response to pooling behavior within each of the five parameter regimes.
\begin{lem}
\label{lem:BRactions}For each regime, $\sigma^{R*}$ on the equilibrium
path is listed in Table \ref{tab:rOptActionPureBetaLe} if $\beta<1-\alpha$
and in Table \ref{tab:rOptActionPureBetaGr} if $\beta>1-\alpha.$
The row labels correspond to the Zero-Dominant (O-D), Zero-Heavy (0-H),
Middle, One-Heavy (1-H), and One-Dominant (1-D) regimes. \end{lem}
\begin{IEEEproof}
See Appendix \ref{ap:brR}.
\end{IEEEproof}
\begin{table}
\caption{\label{tab:rOptActionPureBetaLe}$\sigma^{R*}(1\,|m,e)$ in Pooling
PBNE with $\beta<1-\alpha$}

\begin{centering}
\begin{tabular}{|c|c|c|c|c|}
\hline 
 & $\sigma^{R*}(1\,|\,0,0)$ & $\sigma^{R*}(1\,|\,0,1)$ & $\sigma^{R*}(1\,|\,1,0)$ & $\sigma^{R*}(1\,|\,1,1)$\tabularnewline
\hline 
\hline 
0-D & \multirow{1}{*}{0} & \multirow{1}{*}{0} & \multirow{1}{*}{0} & \multirow{1}{*}{0}\tabularnewline
\hline 
0-H & \multirow{1}{*}{0} & \multirow{1}{*}{1} & \multirow{1}{*}{0} & \multirow{1}{*}{0}\tabularnewline
\hline 
Middle & 0 & 1 & 1 & 0\tabularnewline
\hline 
1-H & \multirow{1}{*}{1} & \multirow{1}{*}{1} & \multirow{1}{*}{1} & \multirow{1}{*}{0}\tabularnewline
\hline 
1-D & \multirow{1}{*}{1} & \multirow{1}{*}{1} & \multirow{1}{*}{1} & \multirow{1}{*}{1}\tabularnewline
\hline 
\end{tabular}
\par\end{centering}

\begin{centering}
\bigskip{}

\par\end{centering}

\begin{centering}
\caption{\label{tab:rOptActionPureBetaGr}$\sigma^{R*}(1\,|m,e)$ in Pooling
PBNE with $\beta>1-\alpha$}
\begin{tabular}{|c|c|c|c|c|}
\hline 
 & $\sigma^{R*}(1\,|\,0,0)$ & $\sigma^{R*}(1\,|\,0,1)$ & $\sigma^{R*}(1\,|\,1,0)$ & $\sigma^{R*}(1\,|\,1,1)$\tabularnewline
\hline 
\hline 
0-D & \multirow{1}{*}{0} & \multirow{1}{*}{0} & \multirow{1}{*}{0} & \multirow{1}{*}{0}\tabularnewline
\hline 
0-H & \multirow{1}{*}{0} & \multirow{1}{*}{0} & \multirow{1}{*}{1} & \multirow{1}{*}{0}\tabularnewline
\hline 
Middle & 0 & 1 & 1 & 0\tabularnewline
\hline 
1-H & \multirow{1}{*}{0} & \multirow{1}{*}{1} & \multirow{1}{*}{1} & \multirow{1}{*}{1}\tabularnewline
\hline 
1-D & \multirow{1}{*}{1} & \multirow{1}{*}{1} & \multirow{1}{*}{1} & \multirow{1}{*}{1}\tabularnewline
\hline 
\end{tabular}
\par\end{centering}

\end{table}

\begin{rem}
\label{rem:dominantRegimes}In the Zero-Dominant and One-Dominant
regimes of all detector classes, $R$ determines $\sigma^{R*}$ based
only on the overwhelming prior probability of one type over the other\footnote{\label{fn:fakeNews}For instance, consider an application to product
reviews in an online marketplace. A product may be low ($\theta=0$)
or high ($\theta=1$) quality. A review may describe the product as
poor ($m=0$) or as good ($m=1$). Based on the wording of the review,
$R$ may be suspicious ($e=1$) that the review is fake, or he may
not be suspicious ($e=0$). He can then buy ($a=1$) or to not buy
($a=0$) the product. According to Remark \ref{rem:dominantRegimes},
if $R$ has a strong prior belief that the product is high quality
($p(1)\approx1$), then he will ignore both the review $m$ and the
evidence $e,$ and he will always buy the product ($a=1$). }. In the Zero-Dominant regime, $R$ chooses $\sigma^{R*}(1\,|\,m,e)=0$
for all $m$ and $e,$ and in the One-Dominant regime, $R$ chooses
$\sigma^{R*}(1\,|\,m,e)=1$ for all $m$ and $e.$ 
\end{rem}

\begin{rem}
\label{rem:nonDominantRegimes}In the Middle regime of both detector
classes, $R$ chooses\footnote{For the same application to online marketplaces as in Footnote \ref{fn:fakeNews},
if $R$ does not have a strong prior belief about the quality of the
product (\emph{e.g.}, $p(1)\approx0.5$), then he will trust the review
(play $a=m$) if $e=0,$ and he will not trust the review (he will
play $a=1-m$) if $e=1.$ } $\sigma^{R*}(1\,|\,m,0)=m$  and $\sigma^{R*}(1\,|\,m,1)=1-m.$ In
other words, $R$ believes the message of $S$ if $e=0$ and does
not believe the message of $S$ if $e=1.$ 
\end{rem}

\subsection{Optimality Condition for $S$\label{sub:Optimality-Condition-for} }

\begin{figure*}
\begin{centering}
\includegraphics[width=1\textwidth]{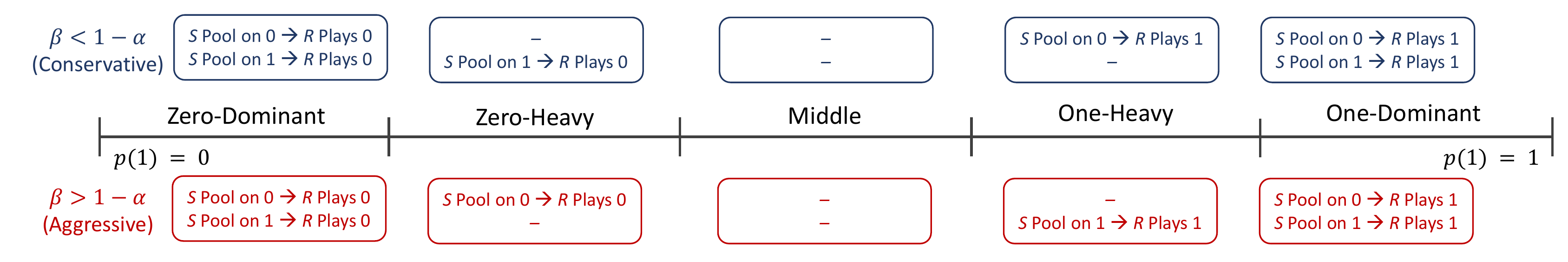}
\par\end{centering}

\caption{\label{fig:eqStratsPure}PBNE in each of the parameter regimes defined
in Fig. \ref{fig:eqRegions}. For $m\in\{0,1\},$ ``$S$ Pool on
$m$'' signifies $\sigma^{S*}(m\,|\,0)=\sigma^{S*}(m\,|\,1)=1.$
For $a\in\{0,1\}$, ``$R$ Plays $a$'' signifies $\sigma^{R*}(a\,|\,0,0)=\sigma^{R*}(a\,|\,0,1)=\sigma^{R*}(a\,|\,1,0)=\sigma^{R*}(a\,|\,1,1)=1.$
Lemma \ref{lem:beliefsRrespSame} gives $\mu^{R}.$ The Dominant regimes
support pooling PBNE on both messages. The Heavy regimes support pooling
PBNE on only one message. The Middle regime does not support any pooling
PBNE.}
\end{figure*}
Next, we must check to see whether each possible pooling strategy
is optimal for $S.$ This depends on what $R$ \emph{would do} if
$S$ were to deviate and send a message off the equilibrium path.
$R$'s action in that case depends on his beliefs for messages off
the path. In PBNE, these beliefs can be set arbitrarily. The challenge
is to see whether beliefs $\mu^{R}$ exist such that each pooling
strategy is optimal for both types of $S.$ Lemmas \ref{lem:beliefsRrespSame}-\ref{lem:beliefsRrespDiffnEER}
give conditions under which such beliefs exist.
\begin{lem}
\label{lem:beliefsRrespSame}Let $m$ be the message on the equilibrium
path. If $\sigma^{R*}(1\,|\,m,0)=\sigma^{R*}(1\,|\,m,1),$ then there
exists a $\mu^{R}$ such that pooling on message $m$ is optimal for
both types of $S.$ For brevity, let $a^{*}\triangleq\sigma^{R*}(1\,|\,m,0)=\sigma^{R*}(1\,|\,m,1).$
Then $\mu^{R}$ is given by,
\[
\forall e\in E,\:\mu^{R}\left(\theta=a^{*}\,|\,1-m,e\right)\geq\frac{\Delta_{1-a^{*}}^{R}}{\Delta_{1-a^{*}}^{R}+\Delta_{a^{*}}^{R}}.
\]

\end{lem}

\begin{lem}
\label{lem:beliefsRrespDiffnEER}If $\sigma^{R*}(1\,|\,m,0)=1-\sigma^{R*}(1\,|\,m,1)$
and $\beta\neq1-\alpha,$ then there does not exist a $\mu^{R}$ such
that pooling on message $m$ is optimal for both types of $S.$\end{lem}
\begin{IEEEproof}
See Appendix \ref{ap:optS} for the proofs of Lemmas \ref{lem:beliefsRrespSame}-\ref{lem:beliefsRrespDiffnEER}.
\end{IEEEproof}
The implications of these lemmas can be seen in the pooling PBNE
results that are presented next.

\subsection{Pooling PBNE\label{sub:Pooling-PBNE}}

Theorem \ref{thm:poolingEq} gives the pooling PBNE of the game.
\begin{thm}
\label{thm:poolingEq}(Pooling PBNE) The pooling PBNE are summarized
by Fig. \ref{fig:eqStratsPure}. \end{thm}
\begin{IEEEproof}
The theorem results from combining Lemmas \ref{lem:BRregions}-\ref{lem:beliefsRrespDiffnEER},
which give the equilibrium $\sigma^{S*},$ $\sigma^{R*},$ and $\mu^{R}.$ \end{IEEEproof}
\begin{rem}
For $\beta<1-\alpha,$ the Zero-Heavy regime admits only a pooling
PBNE on $m=1,$ and the One-Heavy regime admits only a pooling PBNE
on $m=0.$ We call this situation (in which, typically, $m\neq\theta$)
a \emph{falsification convention}. (See Section \ref{sec:compStatics}.)
For $\beta>1-\alpha,$ the Zero-Heavy regime admits only a pooling
PBNE on $m=0,$ and the One-Heavy regime admits only a pooling PBNE
on $m=1.$ We call this situation (in which, typically, $m=\theta$)
a \emph{truth-telling convention}. 
\end{rem}

\begin{rem}
\label{rem:noPoolingMiddle}For $\beta\neq1-\alpha,$ the Middle regime
does not admit any pooling PBNE. This occurs because $R$'s responses
to message $m$ depends on $e,$ \emph{i.e.}, $\sigma^{R*}(1\,|\,0,0)=1-\sigma^{R*}(1\,|\,0,1)$
and $\sigma^{R*}(1\,|\,1,0)=1-\sigma^{R*}(1\,|\,1,1).$ One of the
types of $S$ prefers to deviate to the message off the equilibrium
path. Intuitively, for a conservative detector, $S$ with type $\theta=m$
prefers to deviate to message $1-m,$ because his deception is unlikely
to be detected. On the other hand, for an aggressive detector, $S$
with type $\theta=1-m$ prefers to deviate to message $1-m,$ because
his honesty is likely to produce a false-positive alarm, which will
lead $R$ to guess $a=m.$ Appendix \ref{ap:optS} includes a formal
derivation of this result.
\end{rem}

\subsection{Partially-Separating PBNE\label{sub:Partially-Separating-PBNE}}

For $\beta\neq1-\alpha,$ since the Middle regime does not support
pooling PBNE, we search for partially-separating PBNE. In these PBNE,
$S$ and $R$ play mixed strategies. In mixed-strategy equilibria
in general, each player chooses a mixed strategy that makes the other
players indifferent between the actions that they play with positive
probability. Theorems \ref{thm:partiallySepEqCons}-\ref{thm:partiallySepEqAggr}
give the results.
\begin{thm}
\label{thm:partiallySepEqCons}(Partially-Separating PBNE for Conservative
Detectors) For $\beta<1-\alpha,$ within the Middle Regime, there
exists an equilibrium in which the sender strategies are
\begin{align*}
\sigma^{S*}\left(m=1\,|\,\theta=0\right)= & \frac{\beta^{2}}{\beta^{2}-\alpha^{2}}-\frac{\alpha\beta\Delta_{1}^{R}}{\left(\beta^{2}-\alpha^{2}\right)\Delta_{0}^{R}}\left(\frac{p}{1-p}\right),\\
\sigma^{S*}\left(m=1\,|\,\theta=1\right)= & \frac{\alpha\beta\Delta_{0}^{R}}{\left(\beta^{2}-\alpha^{2}\right)\Delta_{1}^{R}}\left(\frac{1-p}{p}\right)-\frac{\alpha^{2}}{\beta^{2}-\alpha^{2}},
\end{align*}
the receiver strategies are
\begin{align*}
\sigma^{R*}(a=1\,|\,m=0,e=0)=\, & \frac{1-\alpha-\beta}{2-\alpha-\beta},\\
\sigma^{R*}(a=1\,|\,m=0,e=1)=\, & 1,\\
\sigma^{R*}(a=1\,|\,m=1,e=0)=\, & \frac{1}{2-a-b},\\
\sigma^{R*}(a=1\,|\,m=1,e=1)=\, & 0,
\end{align*}
and the beliefs are computed by Bayes' Law in all cases.
\end{thm}

\begin{thm}
\label{thm:partiallySepEqAggr}(Partially-Separating PBNE for Aggressive
Detectors) For any $g\in[0,1],$ let $\bar{g}\triangleq1-g.$ For
$\beta>1-\alpha,$ within the Middle Regime, there exists an equilibrium
in which the sender strategies are
\begin{align*}
\sigma^{S*}\left(m=1\,|\,\theta=0\right)= & \frac{\bar{\alpha}\bar{\beta}\Delta_{1}^{R}}{\left(\bar{\alpha}^{2}-\bar{\beta}^{2}\right)\Delta_{0}^{R}}\left(\frac{p}{1-p}\right)-\frac{\bar{\beta}^{2}}{\bar{\alpha}^{2}-\bar{\beta}^{2}},\\
\sigma^{S*}\left(m=1\,|\,\theta=1\right)= & \frac{\bar{\alpha}^{2}}{\bar{\alpha}^{2}-\bar{\beta}^{2}}-\frac{\bar{\alpha}\bar{\beta}\Delta_{0}^{R}}{\left(\bar{\alpha}^{2}-\bar{\beta}^{2}\right)\Delta_{1}^{R}}\left(\frac{1-p}{p}\right),
\end{align*}
the receiver strategies are
\begin{align*}
\sigma^{R*}(a=1\,|\,m=0,e=0)=\, & 0,\\
\sigma^{R*}(a=1\,|\,m=0,e=1)=\, & \frac{1}{\alpha+\beta},\\
\sigma^{R*}(a=1\,|\,m=1,e=0)=\, & 1,\\
\sigma^{R*}(a=1\,|\,m=1,e=1)=\, & \frac{\alpha+\beta-1}{\alpha+\beta},
\end{align*}
and the beliefs are computed by Bayes' Law in all cases.\end{thm}
\begin{IEEEproof}
See Appendix \ref{ap:partiallySep} for the proofs of Theorems \ref{thm:partiallySepEqCons}-\ref{thm:partiallySepEqAggr}.
\end{IEEEproof}
\begin{figure}
\begin{centering}
\includegraphics[width=0.8\columnwidth]{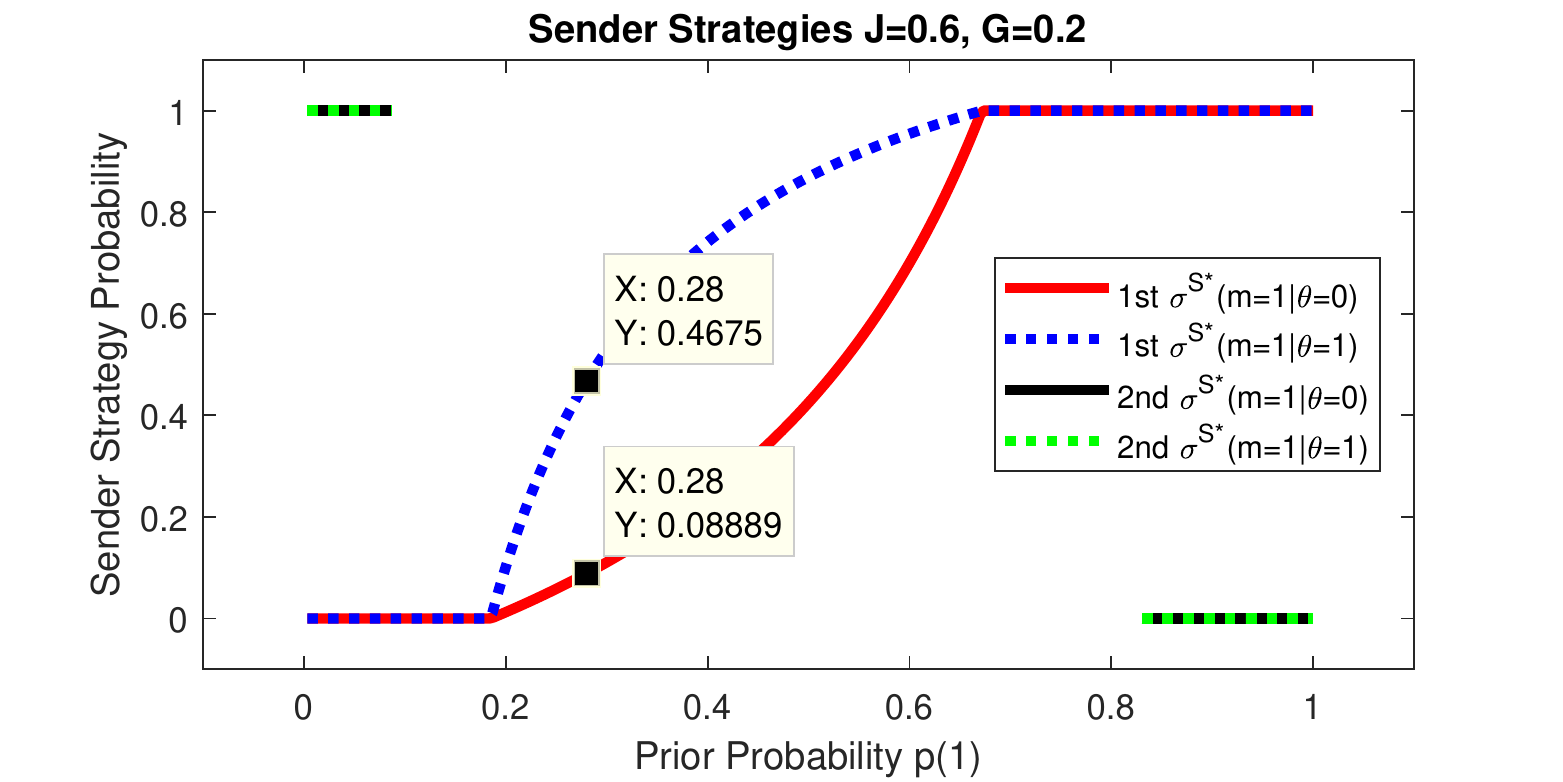}
\par\end{centering}

\begin{centering}
\caption{\label{fig:exampleEqS}Equilibrium sender strategies for $\beta=0.9,$
$\alpha=0.3,$ $\Delta_{0}^{R}=15,$ and $\Delta_{1}^{R}=22.$ The
Dominant regimes of $p(1)$ support both pooling on $m=0$ and $m=1.$
The Heavy regimes ($0.09<p<0.19$ and $0.67<p<0.83$) support only
pooling on $m=0$ and $m=1,$ respectively. The Middle regime does
not support any pooling PBNE, but does support a partially-separating
PBNE.}
\medskip{}

\par\end{centering}

\begin{centering}
\includegraphics[width=0.8\columnwidth]{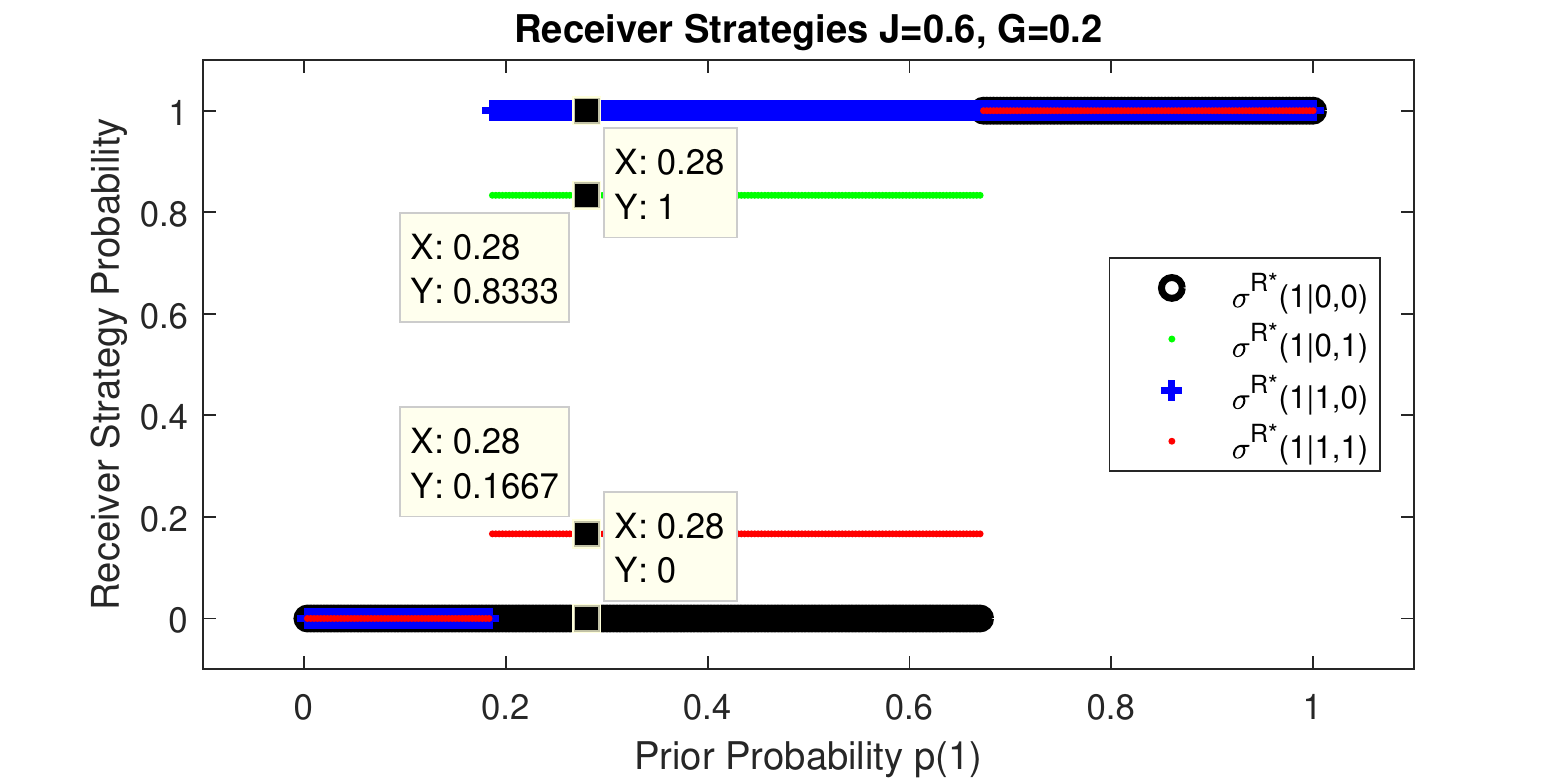}
\par\end{centering}

\caption{\label{fig:exampleEqR}Equilibrium receiver strategies for $\beta=0.9,$
$\alpha=0.3,$ $\Delta_{0}^{R}=15,$ and $\Delta_{1}^{R}=22.$ $R$
plays pure strategies in both the Dominant and Heavy regimes. In the
Middle regime, two strategy components are pure, and two are mixed.}
\end{figure}

\begin{rem}
In Theorem \ref{thm:partiallySepEqCons}, $S$ chooses the $\sigma^{S*}$
that makes $R$ indifferent between $a=0$ and $a=1$ when he observes
the pairs $(m=0,e=0)$ and $(m=1,e=0).$ This allows $R$ to choose
mixed strategies for $\sigma^{R*}(1\,|\,0,0)$ and $\sigma^{R*}(1\,|\,1,0).$
Similarly, $R$ chooses $\sigma^{R*}(1\,|\,0,0)$ and $\sigma^{R*}(1\,|\,1,0)$
that make both types of $S$ indifferent between sending $m=0$ and
$m=1.$ This allows $S$ to choose mixed strategies. A similar pattern
follows in Theorem \ref{thm:partiallySepEqAggr} for $\sigma^{S*},$
$\sigma^{R*}(1\,|\,0,1),$ and $\sigma^{R*}(1\,|\,1,1).$ 
\end{rem}

\begin{rem}
Note that none of the strategies are functions of the sender utility
$u^{S}.$ As shown in Section \ref{sec:compStatics}, this gives the
sender's expected utility a surprising relationship with the properties
of the detector.
\end{rem}
Figure \ref{fig:exampleEqS}-\ref{fig:exampleEqR} depict the equilibrium
strategies for $S$ and $R,$ respectively, for an aggressive detector.
Note that the horizontal axis is the same as the horizontal axis in
Fig. \ref{fig:eqRegions} and Fig. \ref{fig:eqStratsPure}. 

The Zero-Dominant and One-Dominant regimes feature two pooling equilibria.
In Fig. \ref{fig:exampleEqS}, the sender strategies for the first
equilibrium are depicted by the red and blue curves, and the sender
strategies for the second equilibrium are depicted by the green and
black curves. These are pure strategies, because they occur with probabilities
of zero or one. The Zero-Heavy and One-Heavy regimes support only
one pooling equilibria in each case. 

The Middle regime of $p(1)$ features the partially-separating PBNE
given in Theorem \ref{thm:partiallySepEqAggr}. In this regime, Fig.
\ref{fig:exampleEqR} shows that $R$ plays a pure strategy when $e=0$
and a mixed strategy when\footnote{On the other hand, for conservative detectors $R$ plays a pure strategy
when $e=1$ and a mixed strategy when $e=0.$ } $e=1.$ The next section investigates the relationships between these
equilibrium results and the parameters of the game.

\section{Comparative Statics\label{sec:compStatics}}

In this section, we define quantities that we call the \emph{quality}
and \emph{aggressiveness} of the detector. Then we define a quantity
called \emph{truth-induction}, and we examine the variation of truth-induction
with the quality and aggressiveness of the detector.

\subsection{Equilibrium Strategies versus Detector Characteristics}

\begin{figure}
\begin{centering}
\includegraphics[width=0.5\columnwidth]{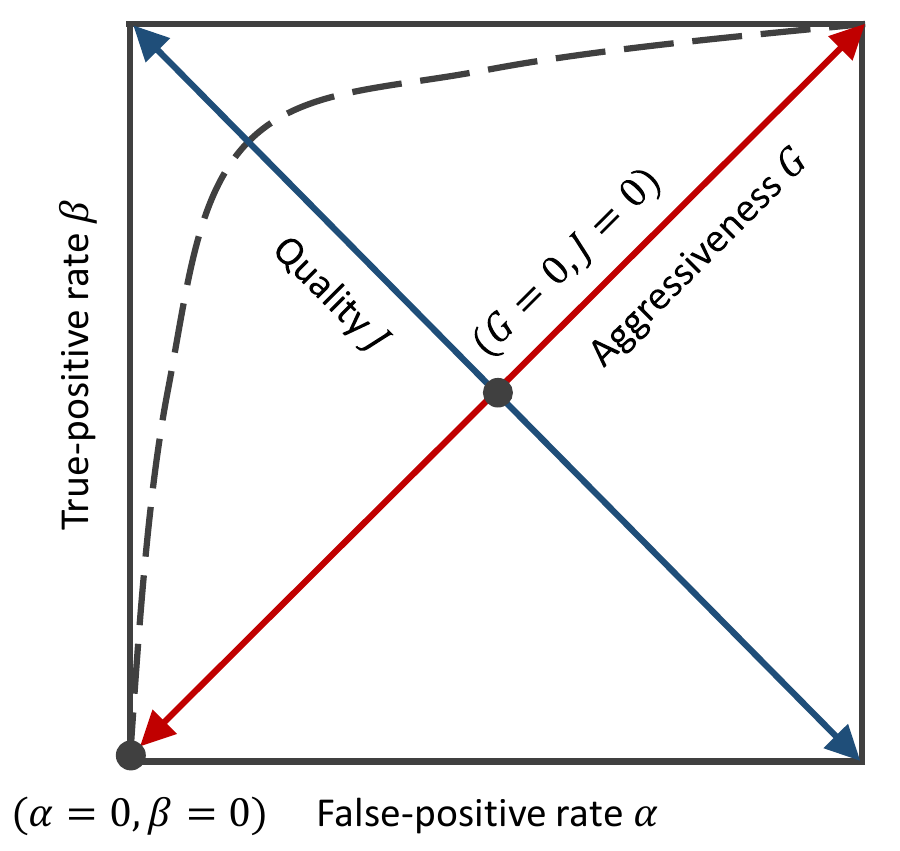}
\par\end{centering}

\caption{\label{fig:axisTransform}The detector characteristics can be plotted
in ROC-space, with $\alpha$ on the horizontal axis and $\beta$ on
the vertical axis. We also parameterize the detector characteristics
by the orthogonal qualities $J\in[-1,1]$ and $G\in[-1,1].$ The dashed
line gives a sample ROC-curve.}
\end{figure}

Consider an alternative parameterization of the detector by the pair
$J$ and $G,$ where $J=\beta-\alpha\in[-1,1],$ and $G=\beta-\left(1-\alpha\right)\in[-1,1].$
$J$ is called \emph{Youden's J Statistic} \cite{youden1950index}.
Since an ideal detector has high $\beta$ and low $\alpha,$ $J$
parameterizes the \emph{quality }of the detector. $G$ parameterizes
the \emph{aggressiveness} of the detector, since an aggressive detector
has $\beta>1-\alpha$ and a conservative detector has $\beta<1-\alpha.$
Figure \ref{fig:axisTransform} depicts the transformation of the
axes. Note that the pair $(J,G)$ fully specifies the pair $(\alpha,\beta).$ 

\begin{figure*}
\begin{centering}
\includegraphics[width=1\textwidth]{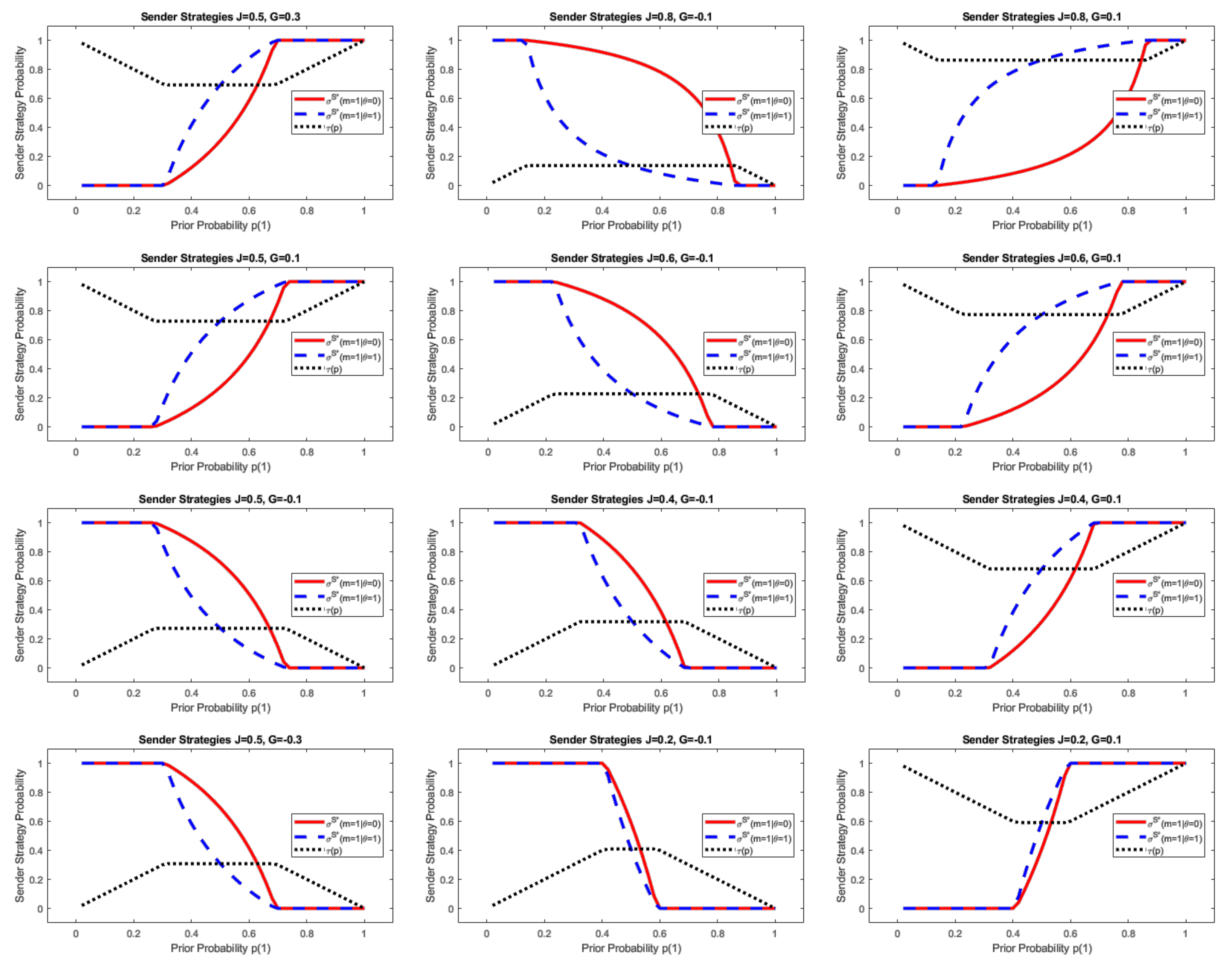}
\par\end{centering}

\caption{\label{fig:multiColSenderStrats}Sender equilibrium strategies $\sigma^{S*}$
and truth-induction rates $\tau.$ Column 1: Detector quality $J$
is fixed and aggressiveness $G$ decreases from top to bottom. Columns
2 and 3: $G$ is fixed and $J$ decreases from top to bottom. Column
2 is for a conservative detector, and Column 3 is for an aggressive
detector. The sender equilibrium strategies are mixed within the Middle
regime, and constant within the other regimes.}
\end{figure*}

Figure \ref{fig:multiColSenderStrats} depicts the influences of $J$
and $G$ on $S$'s equilibrium strategy. The red (solid) curves give
$\sigma^{S*}(m=1\,|\,\theta=0),$ and the blue (dashed) curves represent
$\sigma^{S*}(m=1\,|\,\theta=1).$ Although the Zero-Dominant and One-Dominant
regimes support two pooling equilibria, Fig. \ref{fig:multiColSenderStrats}
only plots one pooling equilibrium for the sake of clarity\footnote{We chose the pooling equilibrium in which $\sigma^{S*}(1\,|\,0)$
and $\sigma^{S*}(1\,|\,1)$ are continuous with the partially-separating
$\sigma^{S*}(1\,|\,0)$ and $\sigma^{S*}(1\,|\,1)$ that are supported
in the Middle regime. }.

In Column 1, $J$ is fixed and $G$ decreases from top to bottom.
The top two plots have $G>0,$ and the bottom two have $G<0.$ There
is a regime change at exactly $G=0.$ At that point, the equilibrium
$\sigma^{S*}(1\,|\,0)$ and $\sigma^{S*}(1\,|\,1)$ flip to their
complements. Here a small perturbation in the characteristics of the
detector leads to a large change in the equilibrium strategies. 

Column $2$ features a conservative detector: $G=-0.1.$ $J$ decreases
from top to bottom. Note that a large $J$ leads to a large Middle
regime, \emph{i.e.}, a large range of $p(1)$ for which $S$ plays
mixed strategies in equilibrium. The detector in Column $3$ is aggressive:
$G=0.1.$ Again, a large $J$ leads to a large Middle regime in which
$S$ plays mixed strategies.

\subsection{Truth-Induction}

Consider the Middle regimes of the games plotted in Columns $2$ and
$3.$ Note that in the Middle regime of Column $2,$ the probabilities
with which both types of $S$ send $m=1$ decrease as $p(1)$ increases,
while in the Middle regime of Column $3,$ the probabilities with
which both types of $S$ send $m=1$ increases as $p(1)$ increases.
This suggests that $S$ is somehow more ``honest'' for the aggressive
detectors in Column $3,$ because $\theta$ and $m$ are more correlated
for aggressive detectors than for conservative detectors. 

In order to formalize this result, let $\sigma^{S*}(m\,|\,\theta;p)$
parameterize the sender's equilibrium strategy by the prior probability
$p\triangleq p(1).$ Then define a mapping $\tau:\,[-1,1]^{2}\times[0,1]\to[0,1],$
such that $\tau(J,G,p)$ gives the \emph{truth-induction rate} of
the detector parameterized by $(J,G)$ at the prior probability\footnote{Feasible detectors have $J\leq1-\left|G\right|.$ In addition, we
only analyze detectors in which $\beta>\alpha,$ which gives $J>0.$ } $p.$ We have
\begin{equation}
\tau\left(J,G,p\right)=\underset{\theta\in\{0,1\}}{\sum}\,p\sigma^{S*}\left(m=\theta\,|\,\theta;p\right).\label{eq:truthIndDef}
\end{equation}
The quantity $\tau$ gives the proportion of messages sent by $S$
for which $m=\theta,$ \emph{i.e.}, for which $S$ tells the truth.
From this definition, we have Theorem \ref{thm:truthInd}.
\begin{thm}
\label{thm:truthInd}(Detectors and Truth Induction Rates) Set $\Delta_{0}^{R}=\Delta_{1}^{R}.$
Then, within prior probability regimes that feature unique PBNE (\emph{i.e.},
the Zero-Heavy, Middle, and One-Heavy Regimes), for all $J\in[0,1]$
and $\forall p\in[0,1],$ we have that
\begin{align*}
\tau\left(J,G,p\right)\leq\frac{1}{2} & \;\;\text{for}\;\;G\in\left(-1,0\right],\\
\tau\left(J,G,p\right)\geq\frac{1}{2} & \;\;\text{for}\;\;G\in\left[0,1\right).
\end{align*}
\end{thm}
\begin{IEEEproof}
See Appendix \ref{ap:truthInd}.\end{IEEEproof}
\begin{rem}
We can summarize Theorem \ref{thm:truthInd} by stating that \emph{aggressive
detectors induce a truth-telling convention}, while \emph{conservative
detectors induce a falsification convention}. 
\end{rem}
The black curves in Fig. \ref{fig:multiColSenderStrats} plot $\tau(p)$
for each of the equilibrium strategies of $S.$  In the regimes with
only one pair of equilibrium strategies for $S,$ $\tau(p)<1/2$ in
Column $2$ and $\tau(p)>1/2$ in Column $3.$ In the Middle regime
of Column $3,$ $\tau(p)$ is largest for detectors with high quality
$J.$

\subsection{Equilibrium Utility}

The \emph{a priori} expected equilibrium utilities of $S$ and $R$
are are the utilities that the players expect \emph{before} $\theta$
is drawn. Denote these utilities by $\tilde{U}^{S}\in\mathbb{R}$
and $\tilde{U}^{R}\in\mathbb{R},$ respectively. For $X\in\{S,R\},$
the utilities are given by
\begin{multline*}
\tilde{U}^{X}=\underset{\theta\in\Theta}{\sum}\,\underset{m\in M}{\sum}\,\underset{e\in E}{\sum}\,\underset{a\in A}{\sum}\,p\left(\theta\right)\\
\sigma^{S*}\left(m\,|\,\theta\right)\lambda\left(e\,|\,\theta,m\right)\sigma^{R*}\left(a\,|\,m,e\right)u^{X}\left(\theta,m,a\right).
\end{multline*}
Based on numerical experiments, we offer two propositions, leaving
formal proofs for future work.
\begin{figure}
\begin{centering}
\includegraphics[width=0.8\columnwidth]{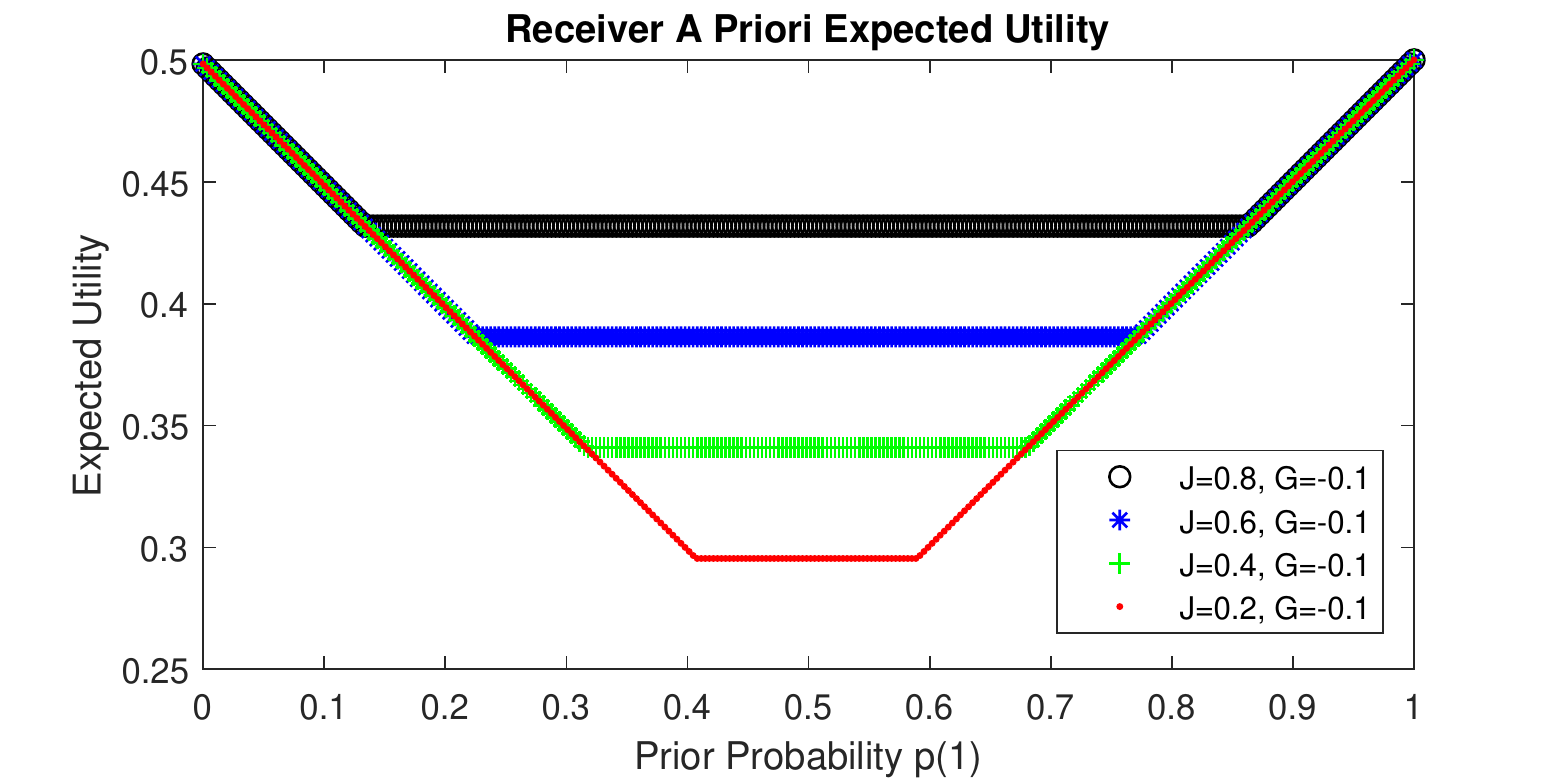}
\par\end{centering}

\caption{\label{fig:utilRwJ}$R$'s \emph{a priori }expected utility for varying
$J.$ Towards the extremes of $p(1),$ $R$'s \emph{a priori} expected
utility does not depend on $J,$ because $R$ ignores $e.$ But in
the middle regime, $R$'s expected utility increases with $J.$ }
\end{figure}

\begin{prop}
\label{prop:utilRwJ}Fix an aggressiveness $G.$ Then, for all $p\in[0,1],$
$\tilde{U}^{R}$ is monotonically non-decreasing in $J.$ 
\end{prop}
Figure \ref{fig:utilRwJ} illustrates Proposition \ref{prop:utilRwJ}.
The proposition claims that $R$'s utility improves as detector quality
improves. In the Middle regime of this example, $R$'s expected utility
increases from $J=0.2$ to $J=0.8.$ Intuitively, as his detection
ability improves, his belief $\mu^{R}$ becomes more certain. In the
Zero-Dominant, Zero-Heavy, One-Heavy, and One-Dominant regimes, $R$'s
equilibrium utility is not affected, because he ignores $e$ and chooses
$a$ based only on prior probability.
\begin{figure}
\begin{centering}
\includegraphics[width=0.8\columnwidth]{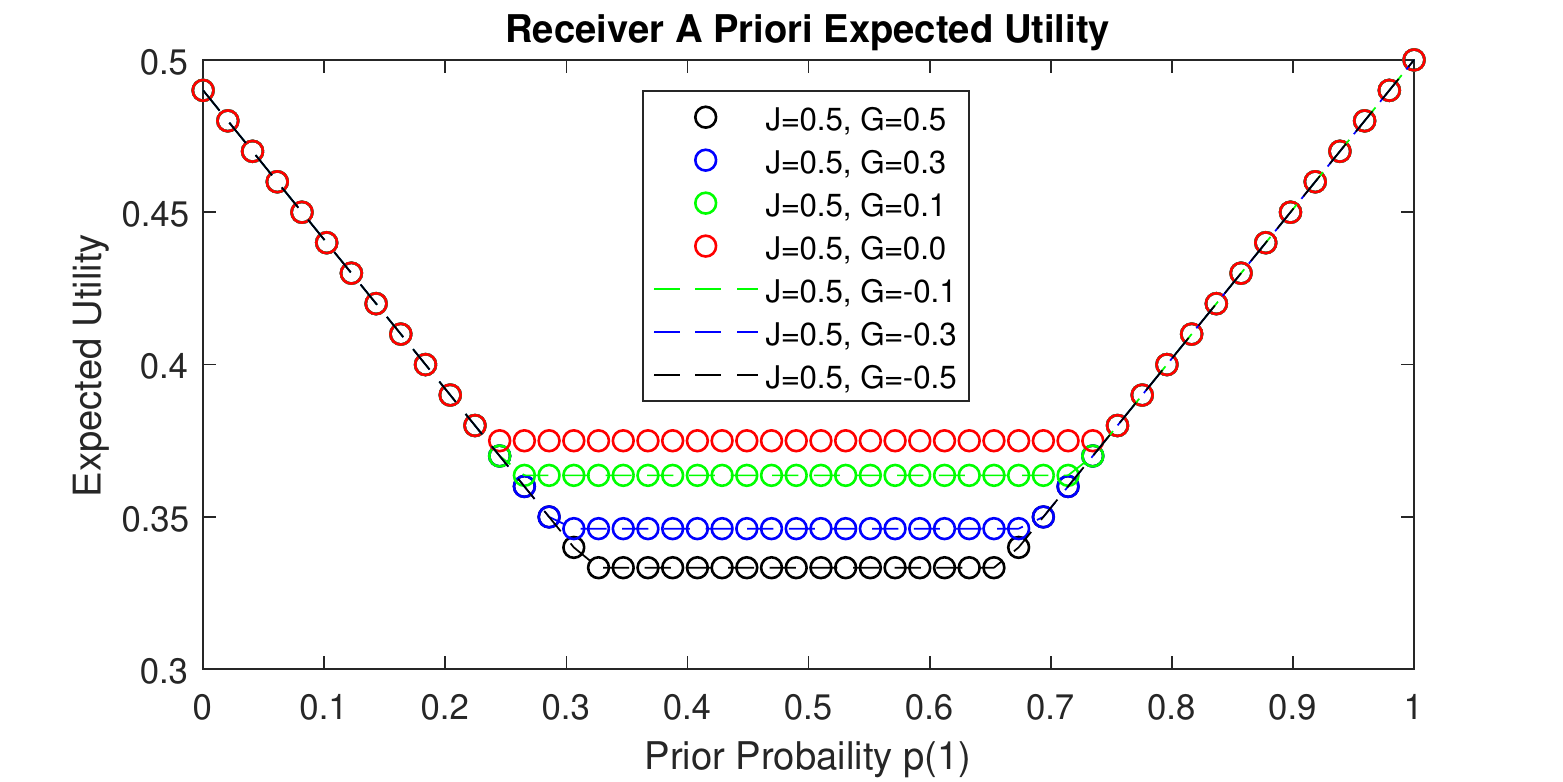}
\par\end{centering}

\caption{\label{fig:utilRwG}$R$'s \emph{a priori }expected utility for varying
$G.$ Towards the extremes of $p(1),$ $R$'s \emph{a priori} expected
utility does not depend on $G,$ because $R$ ignores $e.$ But in
the middle regime, $R$'s expected utility decreases with $|G|.$ }
\end{figure}

\begin{prop}
\label{prop:utilRwG}Fix a detector quality $J.$ Then, for all $p\in[0,1],$
$\tilde{U}^{R}$ is monotonically non-increasing in $|G|.$ 
\end{prop}
For a fixed detector quality, Proposition \ref{prop:utilRwG} suggests
that it is optimal for $R$ to use an EER detector. Figure \ref{fig:utilRwG}
plots $R$'s expected \emph{a priori }equilibrium utility for various
$G$ given a fixed value of $J.$ In the example, the same color is
used for each detector with aggressiveness $G$ and its opposite $-G.$
Detectors with $G\geq0$ are plotted using circles, and detectors
with $G<0$ are plotted using dashed lines. The utilities are the
same for detectors with $G$ and $-G.$ In the Middle regime, expected
utility increases as $|G|$ decreases from $0.5$ to $0.$

\section{Case Study\label{sec:app}}

In this section, we apply signaling games with evidence to the use
of defensive deception for network security. We illustrate the physical
meanings of the action sets and equilibrium outcomes of the game for
this application. We also present several results based on numerical
experiments.

\subsection{Motivation}

Traditional cybersecurity technologies such as firewalls, cryptography,
and role-based access control (RBAC) are insufficient against new
generations of sophisticated and well-resourced adversaries. Attackers
capable of \emph{advanced persistent threats} (\emph{APTs}) use techniques
such as social engineering and hardware exploits to circumvent defenses
and gain insider access. Stealthy and evasive maneuvers are crucial
elements of APTs. Often, attackers are able to remain within a network
for several months or years before they are detected by network defenders
\cite{chen2014study}. This gives the attackers an advantage of information
asymmetry.

To counteract this information asymmetry, defenders have developed
various technologies that detect attackers and provide attackers with
false information. In fact, \emph{honeypots} are classical mechanisms
that achieve both goals. Honeypots are systems placed within a network
in such a manner that the systems are never accessed by legitimate
users. Any activity on the honeypots is evidence that the network
has been breached by an attacker. Sophisticated \emph{research honeypots}
run real services such as a file transfer protocol (FTP) server in
order to allow interaction. In this way, the honeypots gather extensive
information about the attacker's techniques, tools, and procedures
(TTP) \cite{spitzner2003honeynet}.

To some extent, honeypots mimic production systems in order to appear
attractive to attackers. One way of making honeypots appear attractive
is to run programs that simulate user activity. This is a form of
defensive deception. At the same time, attackers may be able to detect
that a deceptive program is running. Hence, honeypot deployment is
the detectable deception that can be captured by our model.

\subsection{Model Description}

\begin{figure}
\centering{}\includegraphics[width=1\columnwidth]{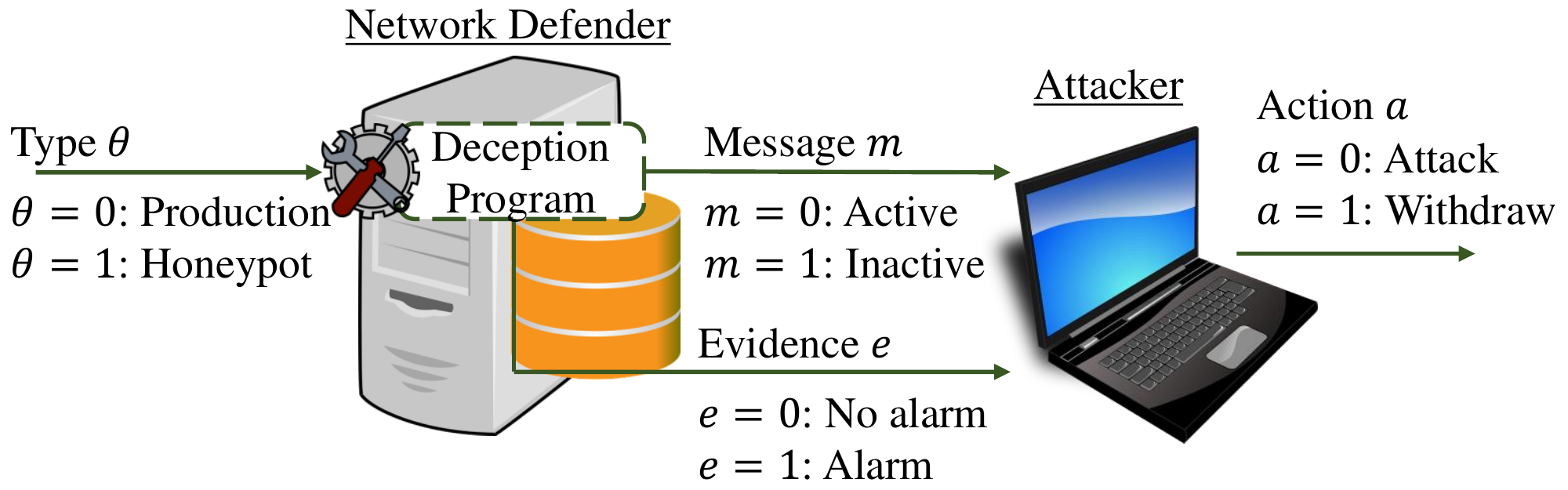}\caption{\label{fig:cybersecurityApp}A defender adds either a production system
or a honeypot to a network. Without any deception, the activity level
of the system is $m=\theta.$ The defender can run a deception program
in order to set $m=1-\theta,$ but the attacker can detect this program
($e=1$) with some probability. Then the defender decides whether
to attack.}
\end{figure}

\subsubsection{Players and Types}

Figure \ref{fig:cybersecurityApp} casts the honeypot interaction
as a signaling game with evidence. The players are the \emph{network
defender} (the sender, $S$) and the \emph{attacker} (the receiver,
$R$). The private information of the sender is the type of the system,
$\theta\in\{0,1\},$ where $\theta=0$ represents a \emph{production
system}, and $\theta=1$ represents a \emph{honeypot}.

\subsubsection{Messages}

Typically, production systems have high activity (since they are accessed
by real users) and honeypots have low activity (since they are not
accessed by real users). Let $m=0$ denote that the system is \emph{active},
and let $m=1$ denote that the system is \emph{inactive}. In order
to deceive $R,$ $S$ can manipulate the activity level of the system.
For instance, with a honeypot, $S$ can send packets from some other
source to the honeypot, create a program to simulate mouse movement
on the honeypot, and create icons on the desktop. All of these make
the honeypot appear active: $m=0.$ Similarly, with a production system,
$S$ can write programs or create user policies that decrease incoming
traffic or limit mouse movement and the number of visible desktop
icons. All of these make the production system appear to be inactive:
$m=1.$

\subsubsection{Evidence}

With some probability, $R$ can detect the use of programs that manipulate
the activity level of the system. For instance, if $S$ makes a honeypot
($\theta=1$) appear active ($m=0$), $R$ may observe that incoming
traffic comes from a single automated source rather than from multiple
human sources. He may also observe that simulated mouse movement follows
patterns different from normal mouse movement, or that desktop icons
are never used. These constitute an \emph{alarm}: $e=1.$ If $R$
does not notice these suspicious signals, then there is \emph{no alarm}:
$e=0.$ Similarly, if $S$ creates programs or user policies for a
production system ($\theta=0$) that limit the incoming traffic and
user activity ($m=1$), $R$ may observe evidence that user behaviors
are being artificially manipulated, which constitutes an alarm: $e=1.$
If $R$ does not notice this manipulation, then there is no alarm:
$e=0.$

\subsubsection{Actions}

After observing the activity level $m$ and leaked evidence $e,$
$R$ chooses an action $a\in\{0,1\}.$ Let $a=0$ denote \emph{attack},
and let $a=1$ denote \emph{withdraw}. Notice that $R$ prefers to
choose $a=\theta,$ while $S$ prefers that $R$ choose $a\neq\theta.$
Hence, Assumptions 2-5 are satisfied. If the cost of running the deceptive
program is negligible, then Assumption 1 is also satisfied\footnote{This is reasonable if the programs are created once and deployed multiple
times.}.

\subsection{Equilibrium Results}

We set the utility functions according to \cite{carroll2011game,pawlick2015deception}.
Consider a detector with a true-positive rate $\beta=0.9$ and a false-positive
rate $\alpha=0.3.$ This detector has $\beta>1-\alpha,$ so it is
an aggressive detector, and the boundaries of the equilibrium regimes
are given in the bottom half of Fig. \ref{fig:eqRegions}. For this
application, the Zero-Dominant and Zero-Heavy regimes can be called
the Production-Dominant and Production-Heavy regimes, since type $\theta=0$
represents a production system. Similarly, the One-Heavy and One-Dominant
regimes can be called the Honeypot-Heavy and Honeypot-Dominant regimes.

We have plotted the equilibrium strategies for these parameters in
Fig. \ref{fig:exampleEqS}-\ref{fig:exampleEqR} in Section \ref{sec:eqResults}.
In the Production-Dominant regime ($p(\theta=1)<0.09$), there are
very few honeypots. In equilibrium, $S$ can set both types of systems
to a high activity level ($m=0$) or set both types of systems to
a low activity level ($m=1$). In both cases, regardless of the evidence
$e,$ $R$ attacks. Next, for the Production-Heavy regime ($0.09<p(1)<0.19$),
the only equilibrium strategy for $S$ is a pooling strategy in which
she leaves production systems ($\theta=0$) active ($m=0$) and runs
a program on honeypots ($\theta=1$) in order to make them appear
active ($m=1$) as well. $R$ is able to detect ($e=1$) the deceptive
program on honeypots with probability $\beta=0.9,$ yet the prior
probability $p(1)$ is low enough that it is optimal for $R$ to ignore
the evidence and attack ($a=1$). 

The Middle regime covers prior probabilities $0.19<p(1)<0.67.$ The
figures display the players' mixed strategies at $p(1)=0.28.$ For
honeypots, $S$'s optimal strategy is to leave the activity level
low with approximately $50\%$ probability ($\sigma^{S*}(1\,|\,1)\approx0.47$)
and to simulate a high activity level with approximately $50\%$ probability.
For production systems, $S$'s optimal strategy is to decrease the
activity level with a low probability ($\sigma^{S*}(1\,|\,0)\approx0.09$)
and to leave the activity level high with the remaining probability. 

In the Middle regime, the receiver plays according to the activity
level---\emph{i.e.}, trusts $S$'s message---if $e=0.$ If $e=1$
when $m=0,$ then most of the time, $R$ does not trust $S$ ($\sigma^{R*}(1\,|\,0,1)\approx0.83$).
Similarly, if $e=1$ when $m=1,$ then most of the time, $R$ does
not trust $S$ ($\sigma^{R*}(1\,|\,1,1)\approx0.17$). The pattern
of equilibria in the Honeypot-Heavy and Honeypot-Dominant regimes
is similar to the pattern of equilibria in the Production-Heavy and
Production-Dominant regimes.

\subsection{Numerical Experiments and Insights}

\subsubsection{Equilibrium Utility of the Sender}

\begin{figure}
\begin{centering}
\includegraphics[width=0.8\columnwidth]{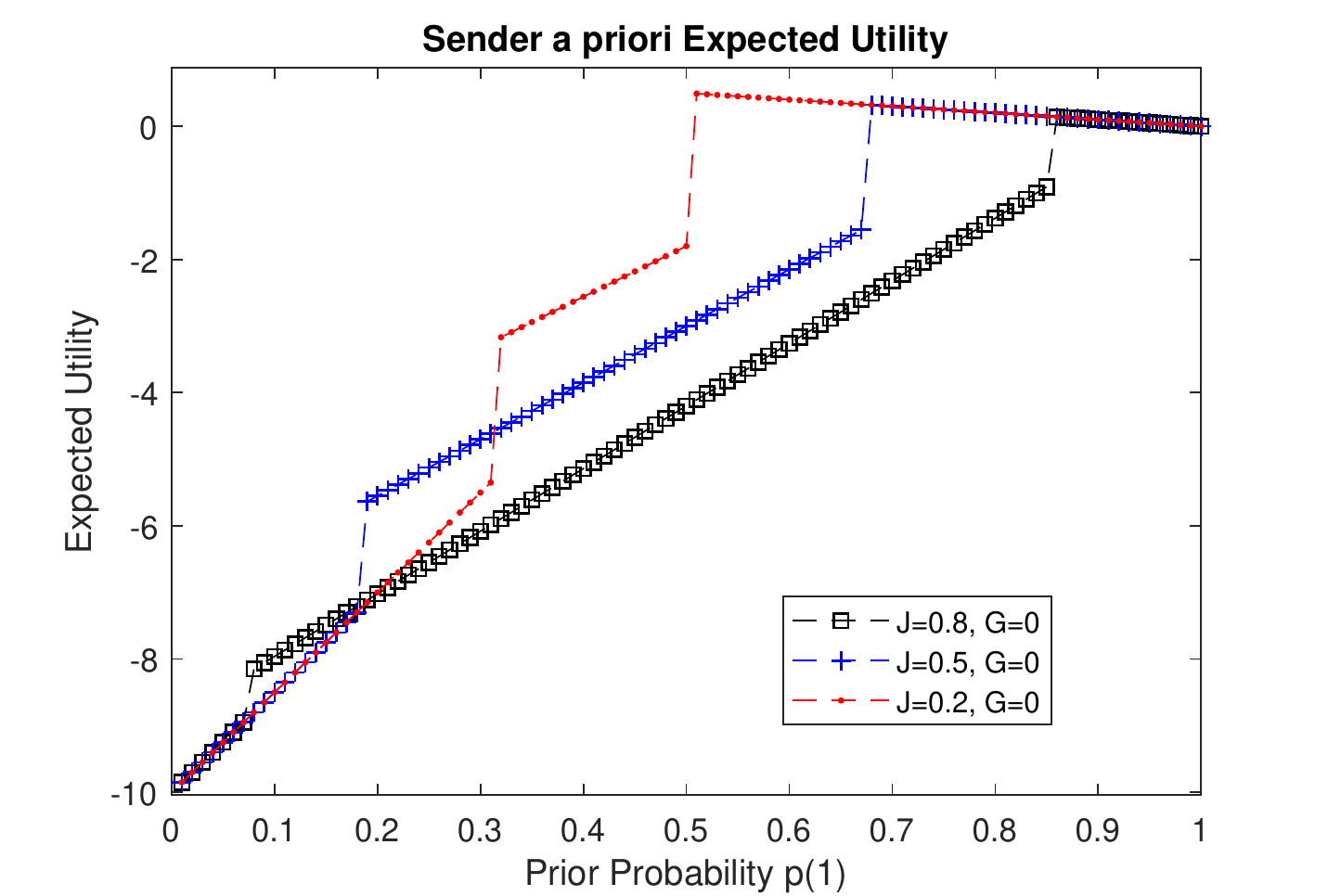}
\par\end{centering}

\begin{centering}
\caption{\label{fig:utilSwJ}$S$'s \emph{a priori }expected utility for varying
$J.$ From least accurate detector to most accurate detector, the
curves are colored red, blue, and black. Suprisingly, for some $p,$
$S$ does better with more accurate detectors than with less accurate
detectors.}
\medskip{}

\par\end{centering}

\begin{centering}
\includegraphics[width=0.8\columnwidth]{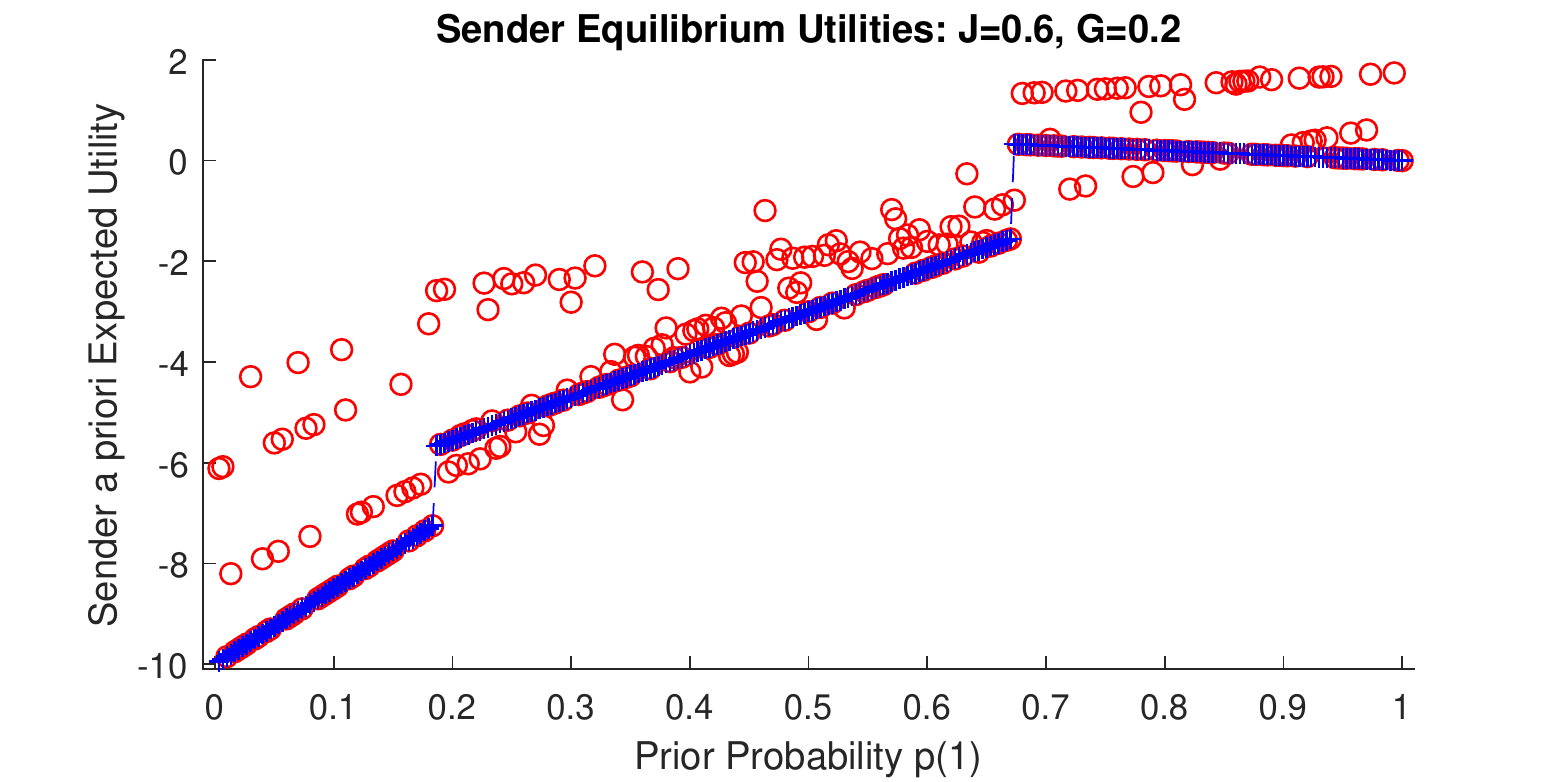}
\par\end{centering}

\caption{\label{fig:senderUtilRsubopt}$S$'s \emph{a priori} expected utility
for playing her equilibrium strategy against 1) $R$'s equilibrium
strategy (in blue crosses), and 2) sub-optimal strategies of $R$
(in red circles). Deviations from $R$'s equilibrium strategy almost
always increase the expected utility of $S.$ }
\end{figure}

Next, Corollary \ref{cor:utilSwJ} considers the relationship between
$S$'s \emph{a priori }equilibrium utility and the quality $J$ of
the detector.
\begin{cor}
\label{cor:utilSwJ}Fix an aggressiveness $G$ and prior probability
$p(1).$ $S$'s \emph{a priori }expected utility $\tilde{U}^{S}$
is not necessarily a decreasing function of the detector quality $J.$ \end{cor}
\begin{IEEEproof}
Figure \ref{fig:utilSwJ} provides a counter-example.
\end{IEEEproof}
Corollary \ref{cor:utilSwJ} is counter-intuitive, because it seems
that the player who attempts deception ($S$) should prefer poor deception
detectors. Surprisingly, this is not always the case. Figure \ref{fig:utilSwJ}
displays the equilibrium utility for $S$ in the honeypot example
for three different $J.$ At $p(1)=0.4,$ $S$ receives the highest
expected utility for the lowest quality deception detector. But at
$p(1)=0.1,$ $S$ receives the highest expected utility for the highest
quality deception detector. In general, this is possible because the
equilibrium regimes and strategies (\emph{e.g.}, in Theorems \ref{thm:poolingEq}-\ref{thm:partiallySepEqAggr})
depend on the utility parameters of $R,$ not $S.$ 

In particular, this occurs because of an asymmetry between the types.
$S$ does very poorly if for a production system ($\theta=0$), she
fails to deceive $R,$ so $R$ attacks ($a=0$). On the other hand,
$S$ does not do very poorly if for a honeypot ($\theta=1$), she
fails to deceive $R,$ who simply withdraws\footnote{For example, consider $p=0.1.$ If $R$ has access to a low-quality
detector, then $p=0.1$ is within the Zero-Heavy regime. Therefore,
$R$ ignores $e$ and always chooses $a=0.$ This is a ``reckless''
strategy that is highly damaging to $S.$ On the other hand, if $R$
has access to a high-quality detector, then $p=0.1$ is within the
Middle regime. In that case, $R$ chooses $a$ based on $e.$ This
``less reckless'' strategy actually improves $S$'s expected utility,
because $R$ chooses $a=0$ less often.} ($a=1$).

\subsubsection{Robustness of the Equilibrium Strategies}

Finally, we investigate the performance of these strategies against
sub-optimal strategies of the other player. Figure \ref{fig:senderUtilRsubopt}
shows the expected utility for $S$ when $R$ acts optimally in blue,
and the expected utility for $S$ when $R$ occasionally acts sub-optimally
in red. In almost all cases, $S$ earns higher expected utility if
$R$ plays sub-optimally than if $R$ plays optimally. 

It can also be shown that $R$'s strategy is robust against a sub-optimal
$S.$ In fact, $R$'s equilibrium utility remains exactly the same
if $S$ plays sub-optimally. 
\begin{cor}
\label{cor:receiverForSsubopt}For all $\sigma^{S}\in\Gamma^{S},$
the \emph{a priori} expected utility of $R$ for a fixed $\sigma^{R*}$
does not vary with $\sigma^{S}.$ \end{cor}
\begin{IEEEproof}
See Appendix \ref{ap:robustProof}.
\end{IEEEproof}
Corollary \ref{cor:receiverForSsubopt} states that $R$'s equilibrium
utility is not affected at all by deviations in the strategy of $S.$
This is a result of the structure of the partially-separating equilibrium.
It suggests that $R$'s equilibrium strategy performs well even if
$S$ is not fully rational.

\section{Discussion\label{sec:discuss}}

We have proposed a holistic and quantitative model of detectable deception
called signaling games with evidence. The detector mechanism can be
conceptualized in two ways. It can be seen as a technology that the
receiver uses to detect deception. For instance, technology for a
GPS receiver can detect spoofed position data based on a lack of variance
in the carrier phase of the signal \cite{psiaki2016attackers}. Alternatively,
the detector can be seen as the inherent tendency of the information
sender to emit cues during deceptive behavior. One example of this
can be found in deceptive opinion spam in online marketplaces; fake
reviews in these marketplaces tend to lack sensorial and concrete
language such as spatial information \cite{ott2011finding}. Of course,
both viewpoints are complementary, because cues of deceptive behavior
are necessary in order for technology to be able to detect deception.

Our equilibrium results include a regime in which the receiver should
chose whether to trust the sender based on the detector evidence (\emph{i.e.},
the Middle regime), and regimes in which the receiver should ignore
the message and evidence and guess the private information using only
the prior probabilities (the Zero-Dominant, Zero-Heavy, One-Heavy,
and One-Dominant regimes). For the sender, our results indicate that
it is optimal to partially reveal the private information in the former
regime, while pooling behavior is optimal in the latter regimes. The
analytical bounds that we have obtained on these regimes can be used
to implement policies online, since they do not require iterative
numerical computation.

We have also presented several contributions that are relevant for
mechanism design. For instance, the mechanism designer can maximize
the receiver utility by choosing an EER detector. Practically, designing
the detector involves setting a threshold within a continuous space
in order to classify an observation as an ``Alarm'' or ``No Alarm.''
For an EER detector, the designer chooses a threshold that obtains
equal false-positive and false-negative error rates.

At the same time, we have shown that aggressive detectors induce a
truth-telling convention, while conservative detectors of the same
quality induce a falsification convention. This is important if truthful
communication is considered to be a design objective in itself. One
area in which this applies is trust management. In both human and
autonomous behavior, an agent is trustworthy if it is open and honest,
if ``its capabilities and limitations {[}are{]} made clear; its style
and content of interactions {[}do{]} not misleadingly suggest that
it is competent where it is really not'' \cite{jones1995human}.
Well-designed detectors can incentivize such truthful revelation of
private information.

Additionally, even deceivers occasionally prefer to reveal some true
information. Our numerical results have shown that the deceiver (the
sender) sometimes receives a higher utility for a high quality detector
than for a low quality detector. This result suggests that cybersecurity
techniques that use defensive deception should not always strive to
eliminate leakage. Sometimes, revealing cues to deception serves as
a deterrent. Finally, the strategies of the sender and receiver are
robust to non-equilibrium actions by the other player. This emerges
from the strong misalignment between their objectives.

Future work could focus on the application of signaling games with
evidence to specific technical domains. These domains often require
adaptations of the model. For instance, problems with continuous type
and message spaces can be addressed by applying a filter that maps
the types and messages into binary spaces (\emph{c.f.}, Section VI
of \cite{pawlick2017trust}). Computational methods can also be used
to address large, discrete action spaces. Finally, signaling games
with evidence can be embedded within larger frameworks in order to
study deception in cyber-physical systems or deception across multiple
links in a network. The present work provides theoretical foundations
and fundamental insights that serve as a foundation for these further
developments.\appendices

\section{Optimal Actions of $R$ in Pooling PBNE\label{ap:brR}}

Consider the case in which both types of $S$ send $m=0.$ On the
equilibrium path, Eq. (\ref{eq:updatePooling}) yields $\mu^{R}(0\,|\,0,0)=(1-\alpha)p(0)/((1-\alpha)p(0)+(1-\beta)p(1))$
and $\mu^{R}(0\,|\,0,1)=\alpha p(0)/(\alpha p(0)+\beta p(1)),$ while
off the equilibrium path, the beliefs can be set arbitrarily. From
Eq. (\ref{eq:receiverOpt}), $R$ chooses action $a=0$\emph{ }(\emph{e.g.},
$R$ believes the signal of $S$) when evidence $e=0$ and $p(0)\geq\Delta_{1}^{R}(1-\beta)/(\Delta_{0}^{R}(1-\alpha)+\Delta_{1}^{R}(1-\beta)),$
 or when evidence $e=1$ and $p(0)\geq\Delta_{1}^{R}\beta/(\Delta_{0}^{R}\alpha+\Delta_{1}^{R}\beta).$

Next, consider the case in which both types of $S$ send $m=1.$ Equation
(\ref{eq:updatePooling}) yields $\mu^{R}(0\,|\,1,0)=(1-\beta)p(0)/((1-\beta)p(0)+(1-\alpha)p(1))$
and $\mu^{R}(0\,|\,1,1)=\beta p(0)/(\beta p(0)+\alpha p(1)),$ which
leads $R$ to choose action $a=1$ (\emph{e.g.} to believe the signal
of $S$) if $e=0$ and $p(0)\leq\Delta_{1}^{R}(1-\alpha)/(\Delta_{0}^{R}(1-\beta)+\Delta_{1}^{R}(1-\alpha)),$
 or if $e=1$ and $p(0)\leq\Delta_{1}^{R}\alpha/(\Delta_{0}^{R}\beta+\Delta_{1}^{R}\alpha).$
The order of these probabilities depends on whether $\beta>1-\alpha.$

\section{Optimal Actions of $S$ in Pooling PBNE\label{ap:optS}}

Let $S$ pool on a message $m.$ Consider the case that $\sigma^{R*}(1\,|\,m,0)=\sigma^{R*}(1\,|\,m,1),$
and let $a^{*}\triangleq\sigma^{R*}(1\,|\,m,0)=\sigma^{R*}(1\,|\,m,1).$
Then $S$ of type $\theta=1-a^{*}$ always successfully deceives $R.$
Clearly, that type of $S$ does not have an incentive to deviate.
But type $S$ of type $\theta=a^{*}$ never deceives $R.$ We must
set the off-equilibrium beliefs such that $S$ of type $\theta=a^{*}$
also would not deceive $R$ if she were to deviate to the other message.
This is the case if $\forall e\in E,$ $\mu^{R}(a^{*}\,|\,1-m,e)\geq\Delta_{1-a^{*}}^{R}/(\Delta_{1-a^{*}}^{R}+\Delta_{a^{*}}^{R}).$
In that case, both types of $S$ meet their optimality conditions,
and we have a pooling PBNE.

But consider the case if $\sigma^{R*}(1\,|\,m,0)=1-\sigma^{R*}(1\,|\,m,1),$
(\emph{i.e.}, if $R$'s response depends on evidence $e$). On the
equilibrium path, $S$ of type $m$ receives utility 
\begin{equation}
u^{S}\left(m,m,m\right)\left(1-\alpha\right)+u^{S}\left(m,m,1-m\right)\alpha,\label{eq:sTypeMeqPath}
\end{equation}
and $S$ of type $1-m$ receives utility 
\begin{equation}
u^{S}\left(1-m,m,1-m\right)\beta+u^{S}\left(1-m,m,m\right)\left(1-\beta\right).\label{eq:sTypeNMeqPath}
\end{equation}
Now we consider $R$'s possible response to messages \emph{off }the
equilibrium path.

First, there cannot be a PBNE if $R$ were to play the same action
with both $e=0$ and $e=1$\emph{ }off the equilibrium path. In that
case, one of the $S$ types could guarantee deception by deviating
to message $1-m.$ Second, there cannot be a PBNE if $R$ were to
play action $a=m$ in response to message $1-m$ with evidence $0$
but action $a=1-m$ in response to message $1-m$ with evidence $1.$
It can be shown that both types of $S$ would deviate. The third possibility
is that $R$ plays action $a=1-m$ in response to message $1-m$ if
$e=0$ but action $a=m$ in response to message $1-m$ if $e=1.$
In that case, for deviating to message $1-m,$ $S$ of type $m$ would
receive utility
\begin{equation}
u^{S}\left(m,1-m,m\right)\beta+u^{S}\left(m,1-m,1-m\right)\left(1-\beta\right),\label{eq:sTypeMoffPath}
\end{equation}
and $S$ of type $1-m$ would receive utility
\begin{multline}
u^{S}\left(1-m,1-m,1-m\right)\left(1-\alpha\right)\\
+u^{S}\left(1-m,1-m,m\right)\alpha.\label{eq:sTypeNMoffPath}
\end{multline}
Combining Eq. (\ref{eq:sTypeMeqPath}-\ref{eq:sTypeMoffPath}), $S$
of type $m$ has an incentive to deviate if $\beta<1-\alpha.$ On
the other hand, combining Eq. (\ref{eq:sTypeNMeqPath}-\ref{eq:sTypeNMoffPath}),
$S$ of type $1-m$ has an incentive to deviate if $\beta>1-\alpha.$
Therefore, if $\beta\neq1-\alpha,$ one type of $S$ always has an
incentive to deviate, and a pooling PBNE is not supported.

\section{\label{ap:partiallySep}Derivation of Partially-Separating Equilibria}

For brevity, define the notation $q\triangleq\sigma^{S*}(1\,|\,0),$
$r\triangleq\sigma^{S*}(1\,|\,1),$ $w\triangleq\sigma^{R*}(1\,|\,0,0),$
$x\triangleq\sigma^{R*}(1\,|\,0,1),$ $y\triangleq\sigma^{R*}(1\,|\,1,0),$
$z\triangleq\sigma^{R*}(1\,|\,1,1),$ and $K\triangleq\Delta_{1}^{R}/(\Delta_{0}^{R}+\Delta_{1}^{R}).$

We prove Theorem \ref{thm:partiallySepEqCons}. First, assume the
receiver's pure strategies $x=1$ and $z=0.$ Second, $R$ must choose
$w$ and $y$ to make both types of $S$ indifferent. This requires
\[
\left[\begin{array}{cc}
\bar{\alpha} & -\bar{\beta}\\
\bar{\beta} & -\bar{\alpha}
\end{array}\right]\left[\begin{array}{c}
w\\
y
\end{array}\right]=\left[\begin{array}{c}
-\alpha\\
-\beta
\end{array}\right],
\]
where $w,y\in[0,1].$ Valid solutions require $\beta\leq1-\alpha.$ 

Third, $S$ must choose $q$ and $r$ to make $R$ indifferent for
$(m,e)=(0,0)$ and $(m,e)=(1,0),$ which are the pairs that pertain
to the strategies $w$ and $y.$ $S$ must satisfy
\[
\left[\begin{array}{cc}
-\bar{\alpha}\bar{p}\bar{K} & \bar{\beta}pK\\
-\bar{\beta}\bar{p}\bar{K} & \bar{\alpha}pK
\end{array}\right]\left[\begin{array}{c}
q\\
r
\end{array}\right]=\left[\begin{array}{c}
-\bar{\alpha}\bar{p}\bar{K}+\bar{\beta}pK\\
0
\end{array}\right].
\]
Valid solutions require $p$ to be within the Middle regime for $\beta\leq1-\alpha.$ 

Fourth, we must verify that the pure strategies $x=1$ and $z=0$
are optimal. This requires 
\[
\frac{\alpha\bar{r}p}{\alpha\bar{r}p+\beta\bar{q}\bar{p}}\leq\bar{K}\leq\frac{\beta rp}{\beta rp+\alpha q\bar{p}}.
\]
It can be shown that, after substitution for $q$ and $r,$ this always
holds. Fifth, the beliefs must be set everywhere according to Bayes'
Law. We have proved Theorem \ref{thm:partiallySepEqCons}.

Now we prove Theorem \ref{thm:partiallySepEqAggr}. First, assume
the receiver's pure strategies $w=0$ and $y=1.$ Second, $R$ must
choose $x$ and $z$ to make both types of $S$ indifferent. This
requires
\[
\left[\begin{array}{cc}
\alpha & -\beta\\
\beta & -\alpha
\end{array}\right]\left[\begin{array}{c}
x\\
z
\end{array}\right]=\left[\begin{array}{c}
\bar{\beta}\\
\bar{\alpha}
\end{array}\right],
\]
where $x,y\in[0,1].$ Valid solutions require $\beta\geq1-\alpha.$ 

Third, $S$ must choose $q$ and $r$ to make $R$ indifferent for
$(m,e)=(0,1)$ and $(m,e)=(1,1),$ which are the pairs that pertain
to the strategies $x$ and $z.$ $S$ must satisfy
\[
\left[\begin{array}{cc}
-\alpha\bar{p}\bar{K} & \beta pK\\
-\beta\bar{p}\bar{K} & \alpha pK
\end{array}\right]\left[\begin{array}{c}
q\\
r
\end{array}\right]=\left[\begin{array}{c}
-\alpha\bar{p}\bar{K}+\beta pK\\
0
\end{array}\right].
\]
Valid solutions require $p$ to be within the Middle regime for $\beta\geq1-\alpha.$ 

Fourth, we must verify that the pure strategies $w=0$ and $y=1$
are optimal. This requires
\[
\frac{\alpha\bar{r}p}{\alpha\bar{r}p+\beta\bar{q}\bar{p}}\leq\bar{K}\leq\frac{\beta rp}{\beta rp+\alpha q\bar{p}}.
\]
It can be shown that, after substitution for $q$ and $r,$ this always
holds. Fifth, the beliefs must be set everywhere according to Bayes'
Law. We have proved Theorem \ref{thm:partiallySepEqAggr}.

\section{Truth-Induction Proof\label{ap:truthInd}}

We prove the theorem in two steps: first for the Middle regime and
second for the Zero-Heavy and One-Heavy regimes.

For conservative detectors in the Middle regime, substituting the
equations of Theorem \ref{thm:partiallySepEqCons} into Eq. (\ref{eq:truthIndDef})
gives
\[
\tau\left(J,G,p\right)=\frac{\bar{\alpha}\bar{\beta}-\bar{\beta}^{2}}{\bar{\alpha}^{2}-\bar{\beta}^{2}}=\frac{1}{2}\left(1-\frac{J}{1-G}\right)\leq\frac{1}{2}.
\]
For aggressive detectors in the Middle regime, substituting the equations
of Theorem \ref{thm:partiallySepEqAggr} into Eq. (\ref{eq:truthIndDef})
gives
\[
\tau\left(J,G,p\right)=\frac{\beta^{2}-\alpha\beta}{\beta^{2}-\alpha^{2}}=\frac{1}{2}\left(1+\frac{J}{1+G}\right)\geq\frac{1}{2}.
\]
This proves the theorem for the Middle regime. 

Now we prove the theorem for the Zero-Heavy and One-Heavy regimes.
Since $\Delta_{0}^{R}=\Delta_{1}^{R},$ all of the Zero-Heavy regime
has $p(1)\leq1/2,$ and all of the One-Heavy regime has $p(1)\geq1/2.$
For conservative detectors in the Zero-Heavy regime, both types of
$S$ transmit $m=1.$ $S$ of type $\theta=0$ are lying, while type
$\theta=1$ are telling the truth. Since $p(1)\leq1/2,$ we have $\tau\leq1/2.$
Similarly, in the One-Heavy regime, both types of $S$ transmit $m=0.$
$S$ of type $\theta=0$ are telling the truth, while $S$ of type
$\theta=1$ are lying. Since $p(1)\geq1/2,$ we have $\tau\leq1/2.$
On the other hand, for aggressive detectors, both types of $S$ transmit
$m=0$ in the Zero-Heavy regime and $m=1$ in the One-Heavy regime.
This yields $\tau\geq1/2$ in both cases. This proves the theorem
for the Zero-Heavy and One-Heavy regimes.

\section{Robustness Proof\label{ap:robustProof}}

Corollary \ref{cor:receiverForSsubopt} is obvious in the pooling
regimes. In those regimes, $\sigma^{R*}(1\,|\,m,e)$ has the same
value for all $m\in M$ and $e\in E$, so if $S$ plays the message
off the equilibrium path, then there is no change in $R$'s action.
In the mixed strategy regimes, using the expressions for $\sigma^{R*}$
from Theorems \ref{thm:partiallySepEqCons}-\ref{thm:partiallySepEqAggr},
it can be shown that, $\forall\theta\in\Theta,$ $m\in M,$ $a\in A,$
\begin{multline*}
\underset{e\in E}{\sum}\,\lambda\left(e\,|\,\theta,m\right)\sigma^{R*}\left(a\,|\,m,0\right)=\\
\underset{e\in E}{\sum}\,\lambda\left(e\,|\,\theta,1-m\right)\sigma^{R*}\left(a\,|\,1-m,0\right).
\end{multline*}
In other words, for both types of $S,$ choosing either message results
in the same probability that $R$ plays each actions. Since $u^{R}$
does not depend on $m,$ both messages result in the same utility
for $R.$ 

\bibliographystyle{plain}
\bibliography{SigEvBib}

\begin{thebibliography}{10}

\bibitem{atzori2010internet}
Luigi Atzori, Antonio Iera, and Giacomo Morabito.
\newblock The {Internet} of things: A survey.
\newblock {\em Comput. Networks}, 54(15):2787--2805, 2010.

\bibitem{avenhaus2002inspection}
Rudolf Avenhaus, Bernhard Von~Stengel, and Shmuel Zamir.
\newblock Inspection games.
\newblock {\em Handbook of Game Theory with Econ. Applications}, 3:1947--1987,
  2002.

\bibitem{carroll2011game}
Thomas~E Carroll and Daniel Grosu.
\newblock A game theoretic investigation of deception in network security.
\newblock {\em Security and Commun. Nets.}, 4(10):1162--1172, 2011.

\bibitem{chen2014study}
Ping Chen, Lieven Desmet, and Christophe Huygens.
\newblock A study on advanced persistent threats.
\newblock In {\em IFIP Intl. Conf. on Communications and Multimedia Security},
  pages 63--72. Springer, 2014.

\bibitem{cott1940adaptive}
Hugh Cott.
\newblock {\em Adaptive Coloration in Animals}.
\newblock Methuen, 1940.

\bibitem{crawford1982strategic}
Vincent~P Crawford and Joel Sobel.
\newblock Strategic information transmission.
\newblock {\em Econometrica: J. of the Econometric Soc.}, pages 1431--1451,
  1982.

\bibitem{fudenberg1991game}
Drew Fudenberg and Jean Tirole.
\newblock {\em Game Theory}.
\newblock The {MIT} Press, 1991.

\bibitem{gneezy2005deception}
Uri Gneezy.
\newblock Deception: The role of consequences.
\newblock {\em American Econ. Review}, 95(1):384--394, 2005.

\bibitem{grossman1981informational}
Sanford~J Grossman.
\newblock The informational role of warranties and private disclosure about
  product quality.
\newblock {\em J. of Law and Econ.}, 24(3):461--483, 1981.

\bibitem{grossman1980disclosure}
Sanford~J Grossman and Oliver~D Hart.
\newblock Disclosure laws and takeover bids.
\newblock {\em J. of Finance}, 35(2):323--334, 1980.

\bibitem{harsanyi1967games}
John~C Harsanyi.
\newblock Games with incomplete information played by ``{Bayesian}'' players.
\newblock {\em Manage. Sci.}, 50(12):1804--1817, 1967.

\bibitem{janczewski2008cyber}
Lech~J. Janczewski and Andrew~M. Colarik.
\newblock {\em Cyber Warfare and Cyber Terrorism}.
\newblock Inform. Sci. Reference, New York, 2008.

\bibitem{jones1995human}
Patricia~M Jones and Christine~M Mitchell.
\newblock Human-computer cooperative problem solving: Theory, design, and
  evaluation of an intelligent associate system.
\newblock {\em IEEE Trans. on Systems, Man, and Cybernetics}, 25(7):1039--1053,
  1995.

\bibitem{kartik2009strategic}
Navin Kartik.
\newblock Strategic communication with lying costs.
\newblock {\em Review of Economic Studies}, 76(4):1359--1395, 2009.

\bibitem{levy2008principles}
Bernard~C Levy.
\newblock Binary and m-ary hypothesis testing.
\newblock In {\em Principles of Signal Detection and Parameter Estimation}.
  Springer Science \& Business Media, 2008.

\bibitem{stanford2016deception}
James~Edwin Mahon.
\newblock The definition of lying and deception.
\newblock In Edward~N. Zalta, editor, {\em Stanford Encyclopedia of
  Philosophy}. Winter 2016 edition, 2016.

\bibitem{milgrom1981good}
Paul~R Milgrom.
\newblock Good news and bad news: Representation theorems and applications.
\newblock {\em Bell J. of Econ.}, pages 380--391, 1981.

\bibitem{minhas2011multifaceted}
Umar~Farooq Minhas, Jie Zhang, Thomas Tran, and Robin Cohen.
\newblock A multifaceted approach to modeling agent trust for effective
  communication in the application of mobile ad hoc vehicular networks.
\newblock {\em IEEE Trans. Systems, Man, and Cybernetics, Part C (Applications
  and Reviews)}, 41(3):407--420, 2011.

\bibitem{nash1950equilibrium}
John~F Nash.
\newblock Equilibrium points in n-person games.
\newblock {\em Proc. Nat. Acad. Sci. USA}, 36(1):48--49, 1950.

\bibitem{ott2011finding}
Myle Ott, Yejin Choi, Claire Cardie, and Jeffrey~T Hancock.
\newblock Finding deceptive opinion spam by any stretch of the imagination.
\newblock In {\em Proc. 49th Annual Meeting Assoc. for Computational
  Linguistics: Human Language Technologies}, pages 309--319, 2011.

\bibitem{pawlick2015flip}
Jeffrey Pawlick, Sadegh Farhang, and Quanyan Zhu.
\newblock Flip the cloud: Cyber-physical signaling games in the presence of
  advanced persistent threats.
\newblock In {\em Decision and Game Theory for Security}, pages 289--308.
  Springer, 2015.

\bibitem{pawlick2015deception}
Jeffrey Pawlick and Quanyan Zhu.
\newblock Deception by design: Evidence-based signaling games for network
  defense.
\newblock In {\em Workshop on the Econ. of Inform. Security}, Delft, The
  Netherlands, 2015.

\bibitem{pawlick2017trust}
Jeffrey Pawlick and Quanyan Zhu.
\newblock Strategic trust in cloud-enabled cyber-physical systems with an
  application to glucose control.
\newblock {\em IEEE Trans. Inform. Forensics and Security}, 12(12), 2017.

\bibitem{psiaki2016attackers}
Mark~L Psiaki, Todd~E Humphreys, and Brian Stauffer.
\newblock Attackers can spoof navigation signals without our knowledge.
  {Here's} how to fight back {GPS} lies.
\newblock {\em IEEE Spectrum}, 53(8):26--53, 2016.

\bibitem{buzzfeed2016fake}
Craig Silverman.
\newblock This analysis shows how viral fake election news stories outperformed
  real news on facebook.
\newblock {BuzzFeed} {News}, [Online]. Available:
  https://www.buzzfeed.com/craigsilverman/.

\bibitem{spitzner2003honeynet}
Lance Spitzner.
\newblock The honeynet project: Trapping the hackers.
\newblock {\em IEEE Security \& Privacy}, 99(2):15--23, 2003.

\bibitem{vrij2008increasing}
Aldert Vrij, Samantha~A Mann, Ronald~P Fisher, Sharon Leal, Rebecca Milne, and
  Ray Bull.
\newblock Increasing cognitive load to facilitate lie detection: The benefit of
  recalling an event in reverse order.
\newblock {\em Law and Human Behavior}, 32(3):253--265, 2008.

\bibitem{youden1950index}
William~J Youden.
\newblock Index for rating diagnostic tests.
\newblock {\em Cancer}, 3(1):32--35, 1950.

\bibitem{zhang2010gpath}
Nan Zhang, Wei Yu, Xinwen Fu, and Sajal~K Das.
\newblock g{P}ath: A game-theoretic path selection algorithm to protect tor's
  anonymity.
\newblock In {\em Decision and Game Theory for Security}, pages 58--71.
  Springer, 2010.

\bibitem{zhu2013game}
Quanyan Zhu and Tamer Ba{\c{s}}ar.
\newblock Game-theoretic approach to feedback-driven multi-stage moving target
  defense.
\newblock In {\em Decision and Game Theory for Security}, pages 246--263.
  Springer, 2013.

\bibitem{zhuang2010modeling}
J.~Zhuang, V.~M. Bier, and O.~Alagoz.
\newblock Modeling secrecy and deception in a multiple-period
  attacker--defender signaling game.
\newblock {\em European J. of Operational Res.}, 203(2):409--418, 2010.

\end{thebibliography}

\end{document}